\documentclass[showpacs,amsmath,amssymb,floatfix,onecolumn,secnumroman]{revtex4}

\usepackage[pdftex]{graphicx}
\usepackage{amssymb,amsfonts,amsmath}
\usepackage{graphicx,amsmath}

\begin{document}

\title{Long range electronic tranport in DNA molecules deposited across a disconnected array of metallic nanoparticles}
\author{A.D. Chepelianskii$^{(a,b)}$ , D Klinov$^{(c)}$ , A Kasumov$^{(b)}$ , S Gu\'eron$^{(b)}$ ,
O Pietrement$^{(d)}$ , S Lyonnais$^{(e)}$ and H Bouchiat$^{(b)}$ \\
(a) Cavendish Laboratory, University of Cambridge, J J Thomson Avenue, Cambridge CB3 OHE, UK \\
(b) LPS, Univ. Paris-Sud, CNRS, UMR 8502, F-91405, Orsay, France\\
(c) Shemyakin-Ovchinnikov Institute of Bioorganic Chemistry, Russian Academy 
of Sciences, Miklukho-Maklaya 16/10, Moscow 117871, Russia \\
(d) UMR 8126 CNRS-IGR-UPS, Institut Gustave-Roussy, 39 rue Camille
Desmoulins, 94805 Villejuif Cedex, France \\
(e) Museum National d’Histoire Naturelle, CNRS, UMR7196, Inserm, U565,
43 rue Cuvier, 75005 Paris, France
}


\begin{abstract} 
We report in detail our experiments on the conduction of $\lambda$ DNA molecules over a wide range of temperature deposited 
across slits in a few nanometers thick platinum film. 
These insulating slits were fabricated using focused ion beam etching and characterized 
extensively using near field and electron microscopy. 
This characterization revealed the presence of metallic Ga nanoparticles inside the slits, 
as a result of the ion etching. After deposition of $\lambda$ DNA molecules, using a protocol that 
we describe in detail, some of the slits became conducting and exhibited superconducting fluctuations at low temperatures. 
We argue that the observed conduction was due to transport along DNA molecules, that interacted 
with the Ga nanoparticles present in the slit. At low temperatures when Ga becomes superconducting, 
induced superconductivity could therefore be observed. These results indicate that minute metallic particles 
can easily transfer charge carriers to attached DNA molecules and provide a possible reconciliation between 
apparently contradictory previous experimental results concerning the length over which DNA molecules can conduct electricity. 
\end{abstract}

\maketitle


DNA is a double stranded molecule, with diameter 2 nanometer and length that 
can vary from a few nanometers to centimeters for mammal DNA.
It is rather stiff compared to other polymers and polyelectrolytes with a persistence length 
of $50$ nanometers.
DNA has a double helix structure, each strand contains four possible bases: 
adenine (abbreviated A), cytosine (C), guanine (G) and thymine (T).
The bases C-G and A-T are paired through hydrogen-bonds and appear on complementary 
strands of the double helix. 
The backbone of DNA is formed by sugar 
and negatively charged phosphate groups. 
In solution, these charged groups 
are surrounded by a cloud of positively charged counter-ions which screens 
part of the negative charge. As a result the effective charge of DNA 
in solution is still negative with average charge density $e / 0.7\;{\rm nm}$ 
instead of $e / 0.34\;{\rm nm}$ for the bare chemical charge. 
($0.7\;{\rm nm}$ is the Bjerrum length in water at room temperature).

The possibility of electronic transport through DNA molecules is motivated 
by the existence of an overlap between $\pi$ orbitals between bases stacked 
along the DNA backbone. The stacking-distance between neighbor basepairs is around $0.34\;{\rm nm}$ \cite{Eley},
close to the distance between atomic planes in graphite. 
Hence the overlap between the molecular orbitals could create delocalized electronic states along the DNA chain. 
The theoretical modeling of electron delocalization along the helix is challenging due to 
the presence of a complicated environment, where sugar, phosphates, water and counterions play 
an important role. Most calculations however agree on the presence of a HOMO (highest occupied molecular orbital)
- LUMO (lowest unoccupied molecular orbital) gap of a few ${\rm eV}$. 
The overlap between HOMO/LUMO orbitals 
localized on neighboring basepairs gives an electronic coupling of the order of $0.1\;{\rm eV}$.
This value must be compared to the ionization potential between adjacent basepairs which is for example $0.6\;{\rm eV}$
between guanine and thymine \cite{Endres}. These values suggest that electronic states are mainly localized on a single basepair. 
However this picture can be strongly modified if the molecule is strongly doped/depleted due to
interaction with the metallic contacts. 

Practical interest in conducting DNA molecules is related to their self-assembly properties
which allow to create nanostructures of a specific shape with a 'bottom-up' approach \cite{Seeman}.
It is now possible to manufacture both two dimensional \cite{Rothemund} and three dimensional structures \cite{YuHe,Andersen}
of well defined shape and chemical properties. It is also conjectured that conduction inside DNA 
may play a role in DNA repair mechanisms, whose efficiency is not well understood yet. 



\begin{figure}[ht]
\centering
\includegraphics[width=0.7 \columnwidth]{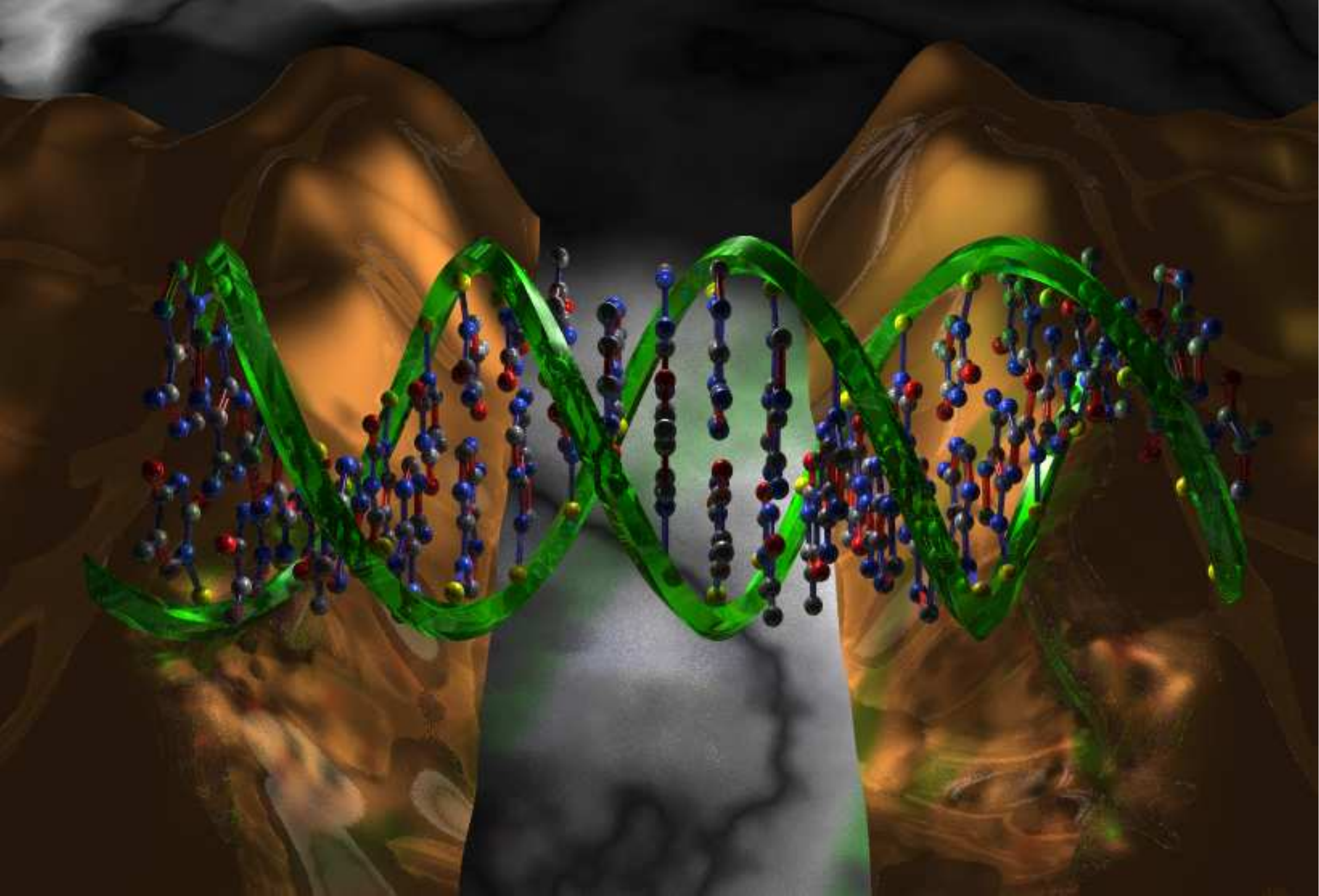}
\caption{Schematic representation of an experiment for the measurement of DNA conduction. 
A small DNA molecule is connected to conducting electrodes separated by an insulating gap. 
}
\label{fig:DNA3D}
\end{figure}

Many experiments were designed to probe transport properties of DNA molecules 
leading to a controversial history that spans across the past decade. 
Conceptually an experiment to measure conduction of a DNA molecule is rather simple 
(see Fig.~\ref{fig:DNA3D}). Nevertheless many contradictory behaviors 
were reported primarily owing to the difficulty of controlling electrode 
fabrication and the interactions between the molecule and its environment 
on a substrate.

Here we will give a brief overview of the experiments in the field
referring to \cite{Endres} for a more thorough review. 
One of the first experiments providing direct evidence of electron transport in DNA molecules 
was reported by Fink and Sch\"onenberger \cite{fink} in 1999. Previous 
spectroscopy experiments also showed charge transfer on distances larger than $4\;{\rm nm}$ \cite{Barton1,Barton2},
however DNA resistivity was not measured directly. 

This experiment was followed by \cite{Porath} where semiconducting behavior was observed 
on poly(G)-poly(C) DNA molecules inserted in a platinum nanogap with separation between electrodes 
around 8 nanometers. The gap reported in \cite{Porath} was in the electron-Volt range. 
The best conduction properties were reported by A. Yu. Kasumov \cite{Kasumov} where conduction was observed at cryogenic 
temperatures. The observation of the superconducting proximity effect suggested that electron transport could be 
coherent over distances larger than $100\;{\rm nm}$. 

Controversy emerged rapidly after the first experiments indicating transport in DNA molecules. 
The experiment \cite{fink} was heavily criticized in Ref.~\cite{Pablo} where conduction 
was attributed to the formation of a carbon contamination layer under electron beam irradiation. 
Absence of transport in DNA on the $100\;{\rm nm}$ scale was reported by several other groups, 
\cite{Storm,zhangcox,Gomez} for DNA on mica and silicon dioxide substrates.
Conduction was probed with DC transport, using gold or platinum electrodes \cite{Storm,zhangcox} or with electric force microscopy \cite{Gomez}.

In view of these results, it appeared that the key ingredient for observation of long range transport in DNA molecules \cite{Kasumov} 
resides in the control of molecule substrate interactions. In the experiment by A. Yu. Kasumov {\it et. al}.  
where conduction could be observed on a 100\;{\rm nm} scale an organic film (pentylamine) was 
deposited onto the substrate before the deposition of DNA and separated molecules from
the insulator/electrode surface \cite{Kasumov,Klinov}.
The key role of the organic film was confirmed in electrostatic mode AFM charge delocalization 
experiments \cite{Heim1} but direct conduction measurements failed at detecting long range conduction \cite{Heim2}.

Recently several experiments observed conduction of short fragments of DNA (with length of a few nanometers).
By repeatedly forming DNA junctions in aqueous buffer solution  \cite{bingqian} concluded 
that resistance was proportional to the molecule length for poly(G)-poly(C) whereas insertion of A-T bases 
led to an exponential decrease of conductance with a decay constant of $0.43 \;{\buildrel _{\circ} \over {\mathrm{A}}}$. 
Using a scanning tunnel microscope operating at cryogenic temperatures \cite{porath} determined 
the excitation spectrum of Poly(C)-Poly(G) molecules which exhibited a clear gap further supporting
the semiconductor model of DNA from Ref.~\cite{Porath}. At last  both 
biological characterization and transport measurements were combined in Ref.~\cite{Barton}
in an experiments where the conduction of a short DNA sequence (15 basepair or $4.5\;{\rm nm}$) 
could be destroyed/restored by introducing/correcting mismatches in the DNA sequence. 
Notwithstanding these recent advances for short molecules, 
the ability of DNA to transport current on length scales of the order of $100\;{\rm nm}$
with rather low resistances around $100\;{\rm k\Omega}$ per molecule is still debated. 

In order to clarify the regime where DNA can transport charge on a relatively long length-scale 
we have tried to reproduce the experiment from Ref.~\cite{Klinov}. 
The description of our experiments will be organized as follow : 
\begin{itemize}

\item {\bf Section~\ref{chap:Dnamica}:} we describe the buffer solutions we use and deposition of $\lambda$-DNA onto mica 

\item {\bf Section~\ref{chap:metal}:} we describe the deposition of $\lambda$ DNA molecules on Pt electrodes,
without the pentylamine organic film. In this experiment molecules are found to be insulating.

\item {\bf Section~\ref{chap:pentylamine}:} we describe deposition of DNA molecules on a substrate 
with pentylamine. We argue that the presence of carbon atoms is necessary to stabilize 
the pentylamine layer and to bind DNA molecules. 

\item {\bf Section~\ref{chap:dnacomb}:} we explain how $\lambda$ DNA molecules can be combed 
across electrodes functionalized with pentylamine. Unfortunately all samples where we 
deposited pentlyamine/molecules ourselves exhibited insulating behavior. 

\item {\bf Section~\ref{chap:FIB}:} gives a description of the fabrication of the insulating gaps 
with a focused ion beam. Starting from this section deposition of DNA molecules was done by D. Klinov
who deposited molecules as in the samples from \cite{Kasumov,Klinov}. Some of the structures became conducting 
after deposition of DNA molecules by D. Klinov. 

\item {\bf Section~\ref{chap:DnaTransport}:} describes transport measurements on the samples 
where conduction was observed down to low temperatures. 

\item {\bf Section~\ref{chap:FibAFMSEM}:} gives an overview of our atomic force microscopy/electron microscopy 
data on the gaps and investigates the connection between conduction and observation of DNA molecules 
across the slits.


\end{itemize}

As we mentioned transport in DNA molecules seems strongly dependent on the molecule 
environment. Unfortunately even if very clean bulk materials can now be synthesized, 
the surface structure and chemistry remain poorly controlled. 
For these reasons technological details become very important and may 
have unforeseen consequences on the surface state of the electrodes. 
Thus in this article we focuss on our understanding of the fabrication process and 
on the control of the surface chemistry of the electrodes, as opposed to our previous 
article \cite{DnaNewJourPhys} where emphasis was made on the interpretation of the low temperature measurements.

\section{Deposition of DNA on a mica substrate} 
\label{chap:Dnamica}

In order to reproducibly deposit DNA molecules on metallic electrodes,
a microscopy technique is needed to observe the molecules on the substrate.
Two main microscopy techniques have sufficient resolution to properly 
image DNA molecules. Transmission electron microscopy (TEM) is the oldest technique 
which allows to study DNA molecules and still offers the best spacial resolution and chemical sensitivity. 
However it has the disadvantage that molecules have to be deposited on thin 
suspended carbon films on a TEM grid. A special surface treatment is required in order to capture 
DNA molecules on hydrophobic carbon films, and often molecules have to be ``stained'' with heavy 
metal salts (uranyl acetate) to improve contrast \cite{Stoeckenius}. Recently atomic force microscopy 
emerged as an alternative technique for visualization of single DNA molecules \cite{Hansma},
in liquid and in air. For this purpose DNA molecules must first be absorbed on a flat surface, 
which is usually mica because it can be easily cleaved in order to provide an atomically flat surface.
In the past few years AFM resolution achieved incredible improvements. For example recently 
it was shown that it is possible to determine the chemical structure of organic molecules absorbed on a surface 
using an atomic force microscope in a mode where individual atoms can be resolved \cite{ScienceAFM,Kawai}.

\begin{figure}[ht]
\centering
\includegraphics[width=0.6 \columnwidth]{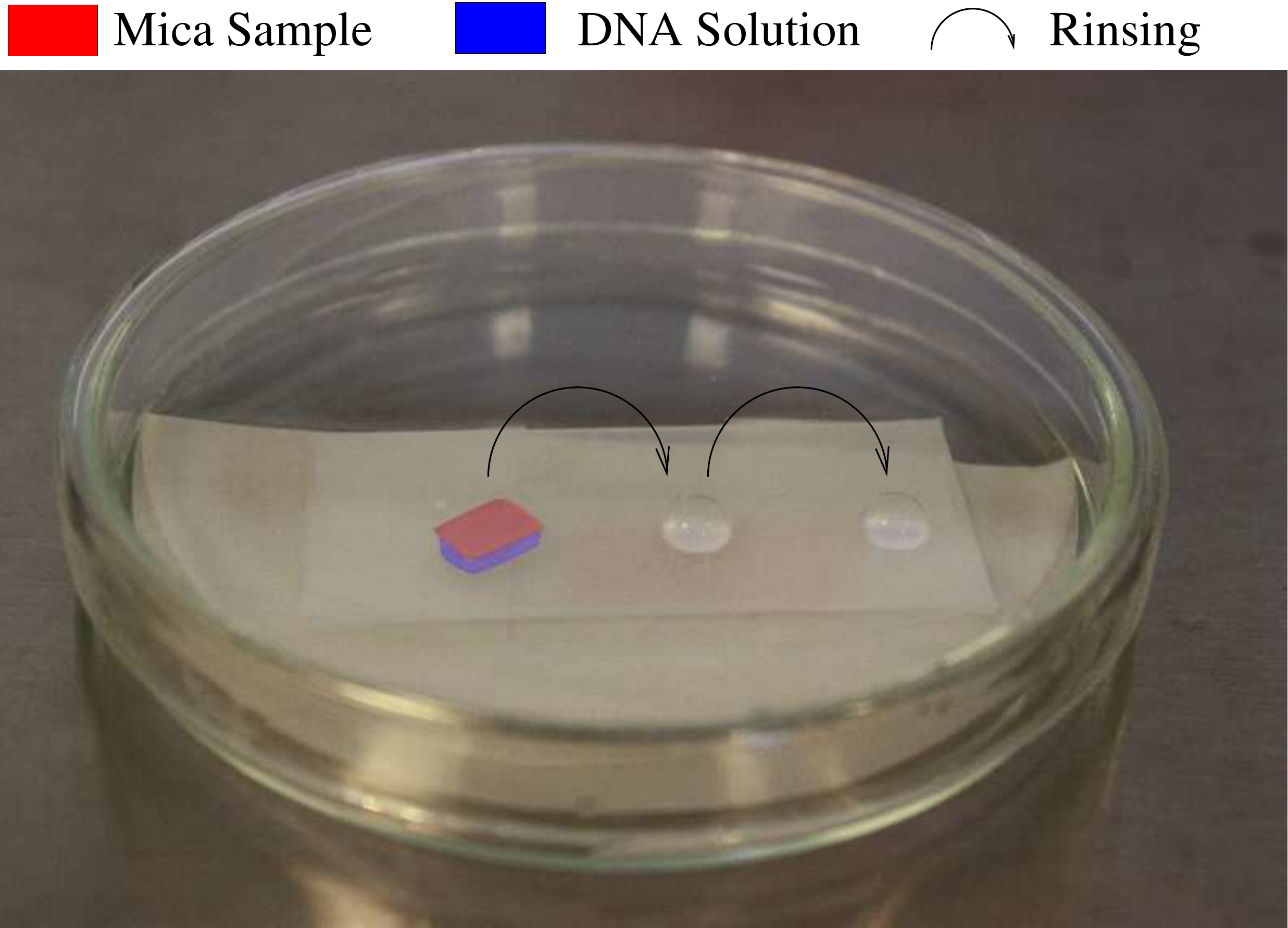}
\caption{Our protocol for deposition of $\lambda$-DNA molecules onto mica. Inside the closed petri box, 
three $100\;{\rm \mu L}$ droplets are deposited on top of a parafilm film: one drop of $\lambda$-DNA, 
and two drops of purified water. The sample is deposited on top of the $\lambda$-DNA drop and 
is subsequently moved onto the water droplets for rinsing. During deposition the petri box is closed 
to ensure stable humidity conditions.
}
\label{fig:ProtocolDima}
\end{figure}

Mica and DNA are both negatively charged in a water solution and therefore there is no adsorption 
of DNA onto Mica when only monovalent salt is present in solution at neutral pH \cite{Pietrement}. 
Binding sites can be created by adding a 
poly-valent salt such as ${\rm Mg Cl_2}$. Table \ref{LambdaDNA} gives the chemical composition of a
a typical solution we used for deposition of lambda DNA molecules onto a mica substrate. 
The choice of the ammonium acetate buffer is unusual, it is chosen mainly for consistency 
with buffers used in \cite{Kasumov,Klinov}. Historically this choice is also motivated by the 
use of ammonium acetate in some standard protocols from electron microscopy on DNA molecules 
\cite{Dykstra}. The solution of Table \ref{LambdaDNA} contains only a very small amount of 
ethylenediaminetetraacetic acid (EDTA) from the native solution, hence it is not suitable 
for DNA storage. EDTA neutralizes metallic ions such as $Ca^{2+}$ or $Fe^{3+}$ 
which are always present in small quantity in solution, and act as 
catalysts for metal-dependent enzymes which can damage DNA. 
For this reason we never used solutions from Tab. \ref{LambdaDNA} for longer than a single day. 

\begin{table}
        \begin{center}
                \begin{tabular}{|c|c|c|}
\hline
Ammonium acetate & ${\rm C H_3 C OO^{-} + N H_4^{+}} $ & 15 mM  \\
\hline
Magnesium chloride & ${\rm Mg Cl_2}  $ & 5 mL \\
\hline 
$\lambda$-DNA  & bought from Invitrogen Cat no. 25250-028  & $5\;{\rm \mu g/ ml}$ \\
\hline 
\hline 
&  Commercial DNA solution contains &  \\
\hline 
$\lambda$-DNA  & 48 502 base pairs & 0.25-0.6 mg/ml \\
\hline
 Tris-HCl      &  ${\rm (HOCH_2)_3 CNH_2 + H Cl}$ (pH 7.4)   & 10 mM \\
\hline
Sodium chloride  & ${\rm Na Cl}$ & 5 mM \\
\hline
EDTA   &     & 0.1 mM \\
\hline
		\end{tabular}
        \caption{Solution for deposition of DNA onto mica for AFM imaging}
        \label{LambdaDNA}        
        \end{center}
\end{table}

\begin{figure}[ht]
\centering
\includegraphics[width=0.7 \columnwidth]{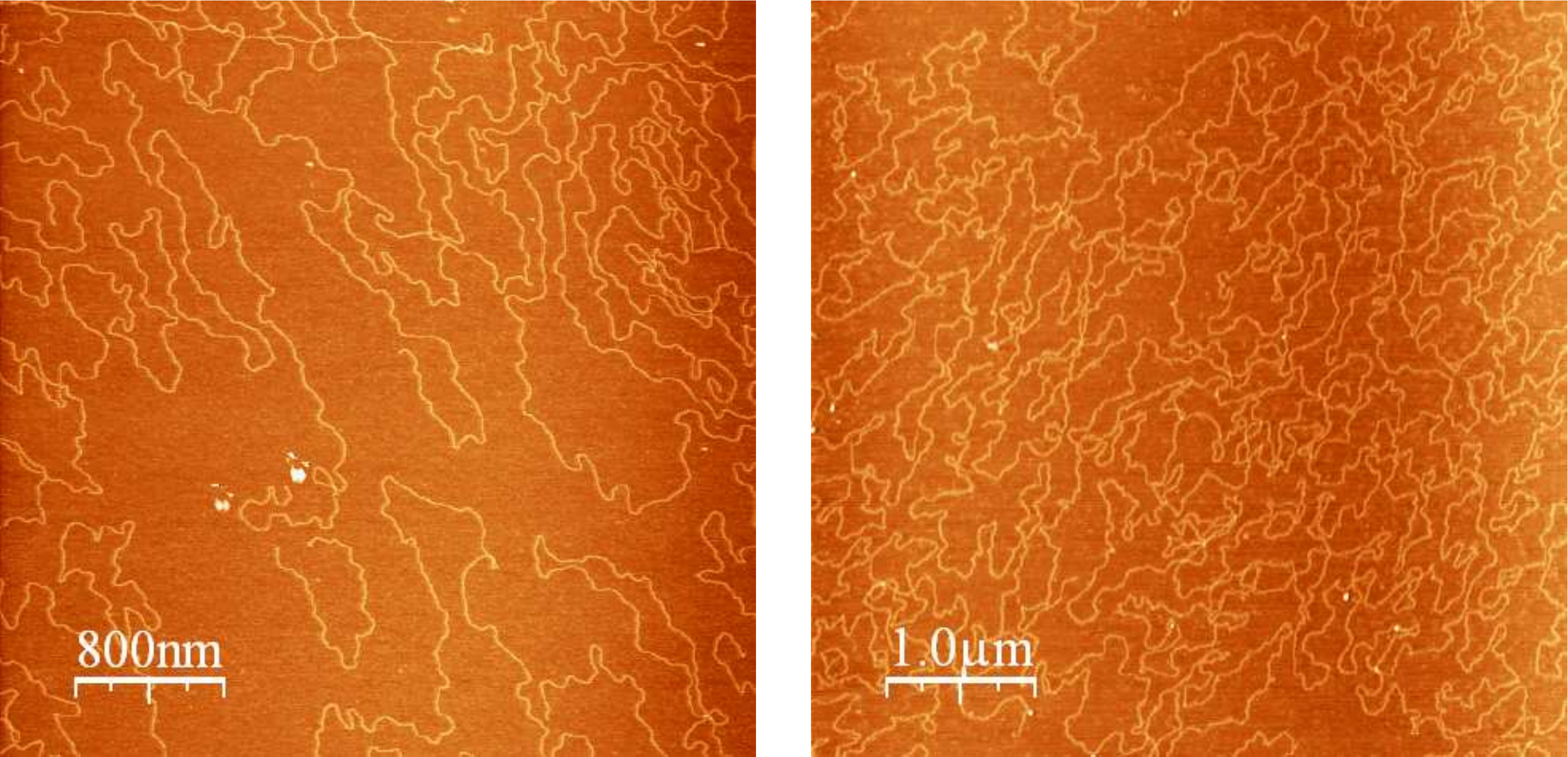}
\caption{AFM image of $\lambda$-DNA molecules absorbed on a mica substrate using solution Tab. \ref{LambdaDNA} and 
following deposition procedure explained in Fig.~\ref{fig:ProtocolDima}. Molecule height is around $0.7$ nm. 
}
\label{fig:DnaAfmDima}
\end{figure}

In order to achieve a reproducible deposition of $\lambda$-DNA onto mica the following protocol 
was suggested to us by D. Klinov. In order to obtain stable humidity conditions during DNA deposition, 
the sample and DNA solution are kept in a closed petri box 
with wet filter paper at the bottom (see Fig.~\ref{fig:ProtocolDima}). 
A drop of DNA solution  (volume $\simeq 100\;{\rm \mu L}$) 
is deposited onto the clean side of a para-film sheet folded around a glass slide.
The freshly cleaved mica sample is then deposited on top of the drop (clean side facing down). 
This reduces the area of the water-air interface during deposition 
and reduces the contamination of the drop. After 10 minutes, many $\lambda$-DNA molecules are attached 
to the mica surface at several anchoring points and the sample can be moved onto of a purified watter 
droplet where salt residues are left to dissolve for around 10 minutes. This rinsing process is repeated two times.

The above protocol allows to achieve reproducible DNA deposition due to several advantages. 
DNA molecules have the time to adsorb on the surface before they undergo the force 
of the meniscus during the drying of the sample, hence they are attached in a state where 
they are not overstretched and keep their natural persistence length. 
Two AFM images from different samples are shown on Fig.~\ref{fig:DnaAfmDima}, 
in both cases AFM shows long molecules undulating on a clean substrate with very 
little contamination. 
This deposition experiment allows us to check that our DNA solution has the right concentration 
and is not contaminated by undesired chemical substances. 

The measured height of DNA is around $0.7$ nm, which contrasts 
with measurements of DNA height on mica in liquid AFM cells where height around $2$ nm can be observed \cite{Hansma}.
Several explanations can be put forward to explain this discrepancy. It is possible that a water 
hydration layer forms near the DNA molecules reducing the apparent height of the molecules. 
Another hypothesis is that DNA may be strongly denaturated by the strong interaction with the surface
when the sample is dry, which can create a transition from B-DNA to A-DNA. 
However we note that there is no significant difference in DNA diameter for A and B forms of DNA \cite{PNASab}. 

For overstretched molecules a transition to the Pauling's P-DNA form is possible and the molecule thickness may indeed be close 
to a nanometer since the phosphate backbones are tightly interwound and the bases are exposed to the 
exterior of the molecule \cite{Croquette}. After deposition of DNA we check that the molecules absorbed on the substrate 
have a persistence length  close to their natural persistence length in solution. This ensures that we 
apply a very limited strain on the molecules and transition to P-DNA form seems unlikely in most of our 
experiments 
(see Fig.~\ref{fig:DnaAfmDima} for undulating molecules, examples of overstretched molecules are shown on Fig.~\ref{fig:DnaPt}).

\section{DNA deposition onto metallic surfaces} 
\label{chap:metal}

For transport measurements DNA must be deposited on a metallic substrate. Many metals (for example Aluminum, Copper, ...) 
can form a thin insulating oxide layer on their surface in atmospheric conditions. While the oxide layer does not influence conduction properties
in the bulk of the metallic film, it can prevent the formation of an electrical contact between DNA 
and the metallic electrodes. Hence the choice of material for electrode is limited 
to noble metals. For experiments on DNA mainly gold, platinum and rhenium have been used so far \cite{Endres} 
although carbon based materials (carbon nanotubes, graphite) emerge as a promising material for contacting DNA 
electrically \cite{Barton}.

\begin{figure}[ht]
\centering
\includegraphics[width=0.7 \columnwidth]{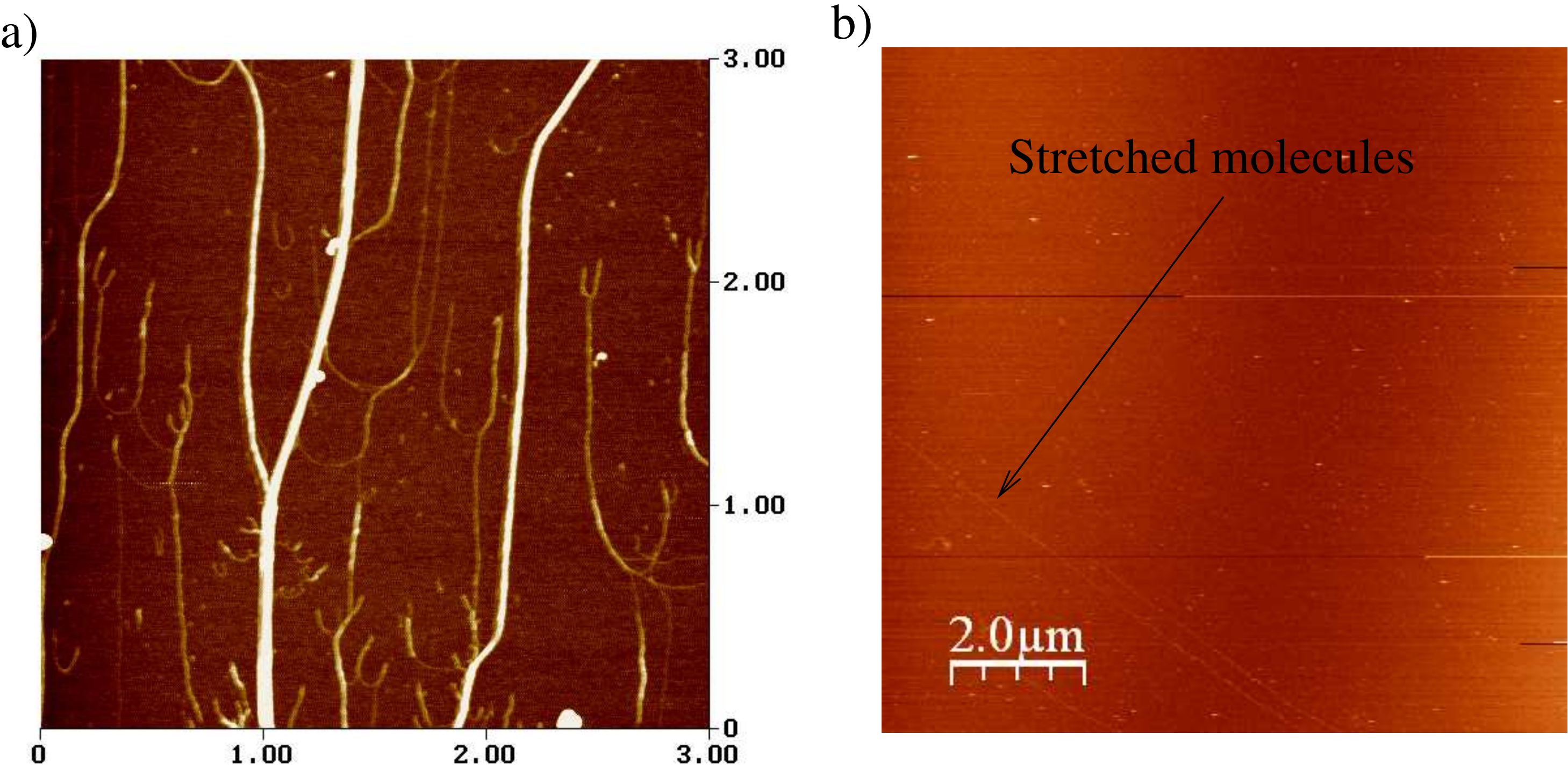}
\caption{Deposition of DNA molecules on a Platinum substrate without treatment of the surface
a) after an incubation time of a few minutes, combing was achieved with a water meniscus 
b) after rinsing under a flow before drying using the protocol described in Section~\ref{chap:dnacomb}.
}
\label{fig:DnaPt}
\end{figure}

In our experiments we have used a thin  Platinum metallic film of a few nanometers (typically between 3 and 5 nm)
deposited by Argon DC-plasma sputtering on a freshly cleaved mica surface. 
Since platinum is deposited everywhere on the sample, no further chemical processing is needed 
and the obtained metallic surface is very clean and chemically inert. As a result there is in principle no 
binding sites to attach DNA molecules to the substrate, and we do not expect DNA adsorption. 
This is not completely true however since DNA molecules have active chemical end groups.
For example it has been proposed that in certain pH ranges DNA ends can expose hydrophobic domains of the bases 
and bind to hydrophic surfaces \cite{pHens}. Hence it is possible that DNA molecules bind to platinum 
through their extremities.

If the DNA solution is incubated a few minutes on the sample both ends of the molecule have in general 
enough time to attach to the substrate. Once the sample is dried the molecules are stretched 
by the water flow leading to a characteristic ``U'' shape of the molecules when the substrate 
is analyzed with an AFM (see Fig.~\ref{fig:DnaPt}.a). As molecules already present on the substrate 
create additional binding sites for the adhesion of other molecules in the solution,
many ropes of DNA molecules can be observed on the Platinum substrate. 
Note that the formation of ropes is less likely if the sample is incubated for a shorter time 
(or rinsed under a continuous flow), in this case (see Fig.~\ref{fig:DnaPt}.b) AFM images 
show only a small number of stretched DNA molecules which are mainly attached through one 
of their extremities.

\begin{figure}[ht]
\centering
\includegraphics[width=0.7 \columnwidth]{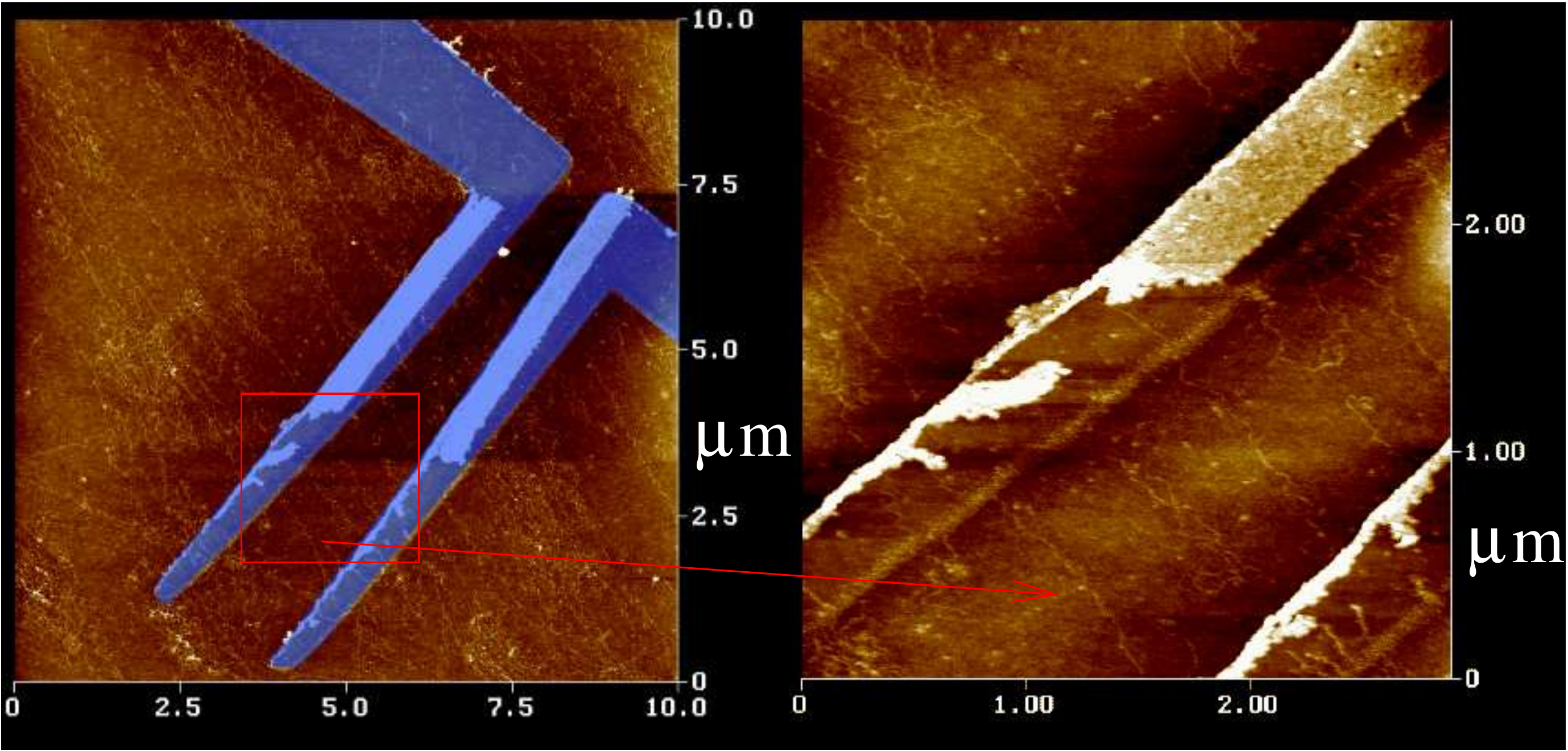}
\caption{AFM image of DNA molecules deposited across Pt electrodes. The electrodes are highlighted in blue on the left image and their height
is approximately $\simeq 3\;{\rm nm}$. They were fabricated with electron beam lithography on a ${\rm Si O_2/Si}$ substrate. 
The right image shows a magnified view of DNA molecules inside the gap. This sample displayed insulating behavior after 
deposition of DNA molecules. (This sample actually underwent a pentylamine plasma treatment, but due to the absence 
of a carbon film on the ${\rm Pt}$ substrate, this treatment was most likely ineffective, see Section \ref{chap:pentylamine}
for a more detailed discussion).
}
\label{fig:DnaPtLitho}
\end{figure}

In the above procedure the DNA molecules were deposited onto the platinum 
film directly after sputtering. Of course transport measurements can not be realized  in 
this configuration. Transport measurements are possible only after electrodes have been 
patterned on the substrate. The simplest way to fabricate an insulating gap is to 
protect the regions where we do not want ${\rm Pt}$ deposition by a MMA/PMMA resist 
that can be patterned beforehand using usual electron-beam lithography techniques. After sputtering 
the resist is dissolved in hot acetone and a gap is formed. The presence of organic residues 
originating from imperfect removal of the PMMA film changes the adsorption properties 
of DNA molecules on the substrate. 
On this ``contaminated'' substrate it becomes possible to deposit molecules without over-stretching 
or creation of ropes (see Fig.~\ref{fig:DnaPtLitho}) although the success rate  is 
small since the surface state of these samples is rather poorly controlled
(indeed in many cases DNA molecules do not bind at all to sample).
These sample showed insulating behavior even if several DNA molecules 
crossed the gap between the ${\rm Pt}$ electrodes (see for e.g. Fig.~\ref{fig:DnaPtLitho}, shortest distance 
between electrodes was around $500\;{\rm nm}$ in this sample). 
This insulating behavior is consistent with the experimental findings 
from \cite{Storm}, which indicated that DNA is an insulator when it is deposited on Silicon/and mica surfaces. 

Because of the poor reproducibility of DNA deposition on bare metallic samples,
and confirmed absence of conduction we subsequently focused 
onto deposition of DNA on metallic electrodes functionalized by a pentylamine plasma
as proposed in Refs. \cite{Kasumov,Klinov} where conduction on samples with DNA was observed at low temperature.

\section{Pentylamine plasma functionalization for deposition of DNA molecules}
\label{chap:pentylamine}

\begin{figure}[ht]
\centering
\includegraphics[width=0.5 \columnwidth]{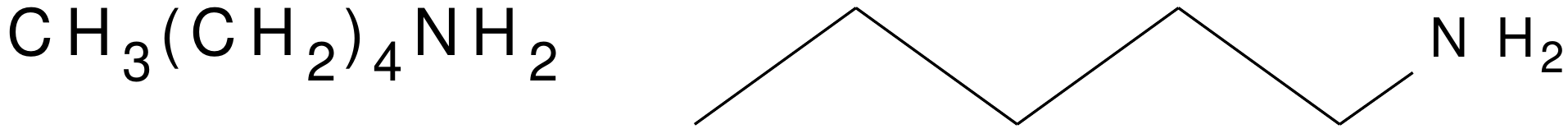}
\caption{Chemical structure of the pentylamine molecule 
}
\label{fig:Penty}
\end{figure}

The use of the pentylamine molecule (see Fig.~\ref{fig:Penty})  was introduced by Dubochet \cite{Dubochet} as a mean to 
render carbon support films for electron microscopy hydrophilic in order to make possible adsorption 
of biological molecules. The sample is commonly placed between two parallel electrode plates 
in a vacuum chamber with weak pentylamine pressure ($2.5 \;{\rm millibar}$). A high voltage ($80\;{\rm V}$) is then applied to the plates 
creating a glow discharge. The ionic species produced during the discharge are deposited 
on the substrate and create a positively charged background favorable for DNA adsorption (typical 
discharge time was $30\;{\rm s}$ in our experiments, reported pressure and voltage values correspond to 
those used for the setup in the group ``Microscopie mol\'eculaire'' at Institut Gustave Roussy).

The chemical structure of the deposited chemical species is not well characterized and is certainly complex
(see also Table \ref{tab:Chem}).
For example it is known that a discharge of an ${\rm CF_4 / O_2}$ gas mixture, creates 
${\rm CF_3^+, CF_2^+, O_2^+, O^-, F^-}$ ions and  ${\rm C F_3, CF_2, O, F}$  radicals \cite{Liberman}
Note however that the fraction of dissociated species among the gas molecules is usually very small (of the order of $10^{-5}$) 
in this discharges used for material processing, hence it is highly probable that not only dissociated 
molecules are deposited on the substrate. 

\begin{table}
        \begin{center}
                \begin{tabular}{|c|c|}
\hline
		  Bond dissociation energy & kJ/mol \\
\hline
		  ${\rm H-NH_2}$ & 450  \\
		  ${\rm H-CH_3}$ & 439  \\
		  ${\rm CH_3-CH_3}$ & 376  \\
		  ${\rm C_2H_5-CH_2NH_2}$ & 336  \\

		  ${\rm C_6H_5CH_2-NH_2}$ & 297  \\
\hline
		  Ionization energy &  kJ/mol  \\
\hline
		  ${\rm H}$ & 1312 \\

		  ${\rm CH_3NH_2}$ & 826 \\

		  ${\rm C_5 H_{11} N}$ & 726 \\
\hline
		\end{tabular}
        \caption{Dissociation energies for bonds present in the pentylamine molecule, 
	    and ionization energies of methylamine (${\rm CH_3NH_2}$) and piperidine (${\rm C_5 H_{11} N}$) \cite{Lide}.
	    The last two molecule are chemically close to pentylamine for which data is not available.
	    Since the bond dissociation energies are all comparable and smaller than the typical ionization energies it seems very likely that all possible 
	    chemical species are present in the plasma. }
        \label{tab:Chem}
        \end{center}
\end{table}

The pentylamine discharge technique was adapted  to attach DNA molecules to conducting electrodes in the experiments 
from Refs. \cite{Kasumov,Klinov}. 
It constitutes the main difference with other studies where in most cases the substrate was silicon dioxide.
Hence we dedicated considerable efforts to identify the substrate on which this treatment yields effective 
binding of DNA molecules. These studies led us to the conclusion that reproducible adsorption of DNA with pentylamine treatment 
occurs only on carbon coated substrates, 
which are similar to the carbon support films for electron microscopy. 
In this respect the analysis in \cite{Klinov} is somewhat misleading since it claims that pentylamine 
can form a polymer film on mica suitable for adsorption of DNA. Below we summarize the results of our 
DNA deposition experiments on different substrates using the pentylamine technique. In all cases the DNA deposition was 
attempted rapidly (at most one hour) after the glow discharge since we have observed that the efficiency of the pentylamine layer
at binding DNA molecules decreases quickly once it is exposed to ambient air.

\begin{figure}[ht]
\centering
\includegraphics[width=0.7 \columnwidth]{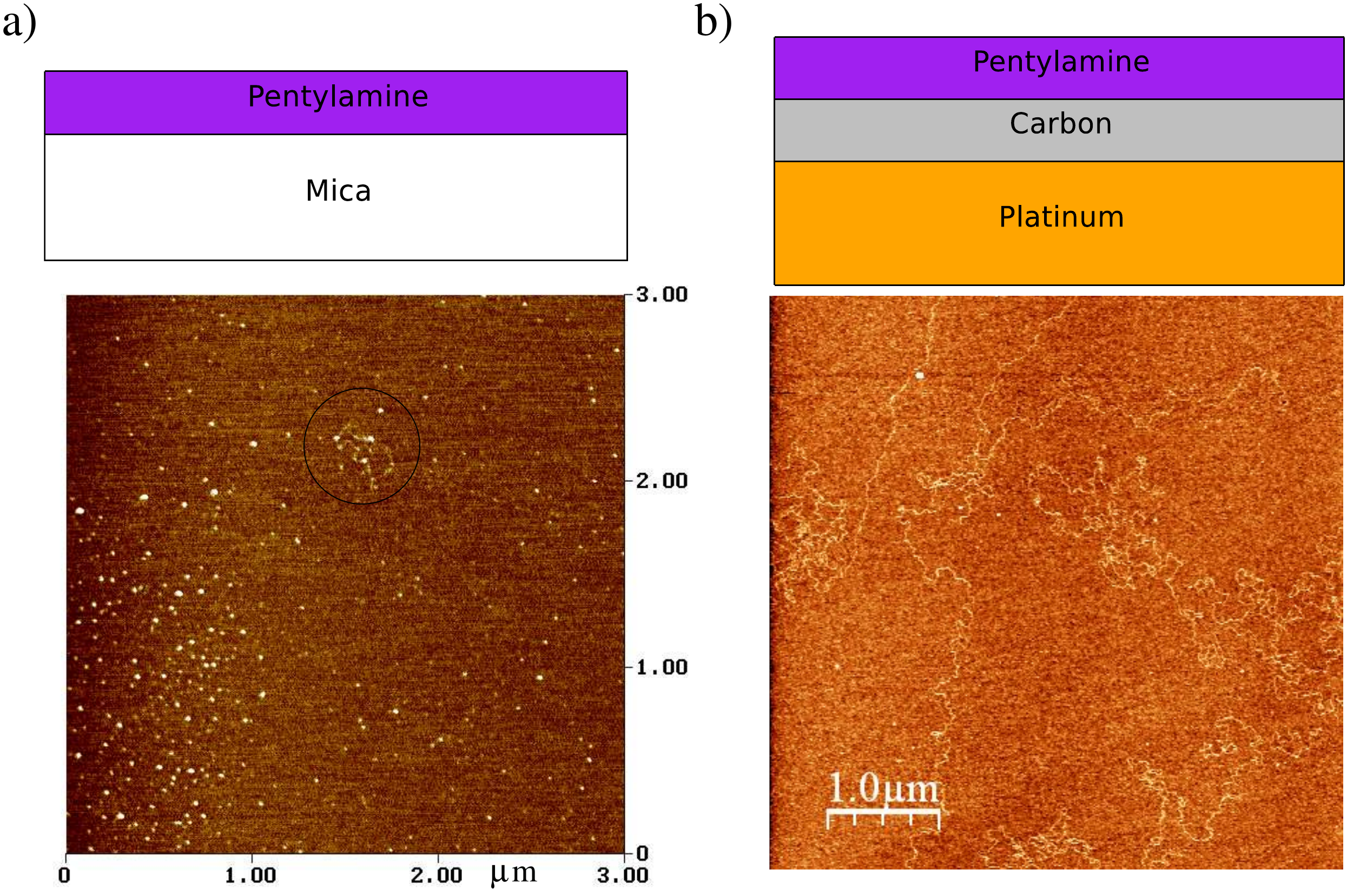}
\caption{a) AFM image of a mica substrate functionalized with pentylamine after an attempt to deposit DNA 
molecules. b) AFM image of a Platinum/Carbon bilayer functionalized with pentylamine after deposition of DNA 
molecules. 
}
\label{fig:MiPenty}
\end{figure}

Our attempts to deposit DNA on a freshly cleaved mica substrate functionalized by pentylamine 
plasma were never successful. An example AFM image of the mica substrate after  pentylamine plasma and DNA deposition 
is shown on Fig.~\ref{fig:MiPenty}. Only a single feature reassembling a DNA molecule can be distinguished inside the  $3\;{\rm \mu m} \times 3\;{\rm \mu m}$ 
scan area : this is much less than the number of molecules on the Figs.~\ref{fig:DnaAfmDima},\ref{fig:DnaPtLitho} which 
have similar scan size. The absence of molecules on the mica substrate is puzzling since 
DNA binds to both pentylamine functionalized surfaces and bare mica. 
A possible explanation is that in a first stage molecules bind 
to the pentylamine film on top of the mica surface. 
In a second stage this film is washed away from the surface when the sample is dried
ripping away the DNA molecules on top of the pentylamine film. 
Indeed the pentylamine film must be destroyed at some stage of the deposition 
because the density of DNA molecules on the surface is extremely small compared to the densities
achieved on the Pt/C surface with the same functionalization. 
It can not be destroyed immediately when the drop is deposited on the surface, 
otherwise DNA molecules would bind to the exposed mica surface. 
To summarize in the above scenario pentylamine masks the mica surface from the DNA solution and 
is at least partially removed when the sample is dried, carrying away the attached DNA molecules
(this assumes that the layer formed during the plasma discharge is continuous).

Deposition of DNA molecules on a platinum film coated with an evaporated layer of amorphous 
carbon (around 10 nm) were reproducible and successful (see Fig.~\ref{fig:MiPenty}). Most likely the free radicals 
created during the glow discharge react with the amorphous carbon on the surface, and anchor the pentylamine layer. 
As seen in this image the molecules are not overstretched, it may also seem that their 
persistence length is smaller than on the mica substrate (see Fig.~\ref{fig:DnaAfmDima}).
This observation is supported by other quantitative studies of DNA adsorption 
on positively charged surfaces \cite{DnaPositive}. Following Ref.~\cite{Klinov} we have 
also tried to deposit DNA with the pentylamine treatment directly onto platinum without 
the carbon layer. For clean platinum films only a low density of overstretched molecules 
could be detected on the surface after DNA deposition. This most likely indicates 
that the pentylamine is removed during the DNA deposition and only a few molecules 
bind to the platinum surface through their ends with a mechanism similar to that 
described in Section \ref{chap:metal}. After discussions with D. Klinov, we found
that the Platinum deposited in \cite{Klinov} actually contained a certain amount of carbon 
(around 10\%) which allowed fixation of the pentylamine.

In conclusion the adhesion of the pentylamine is reliable only on surfaces with a high enough density 
of carbon atoms that can bind with the ions/radicals produced during the glow discharge 
sticking the pentylamine to the surface. In this case the deposition of DNA molecules is reproducible 
and molecules are not overstretched. 
The role of the carbon atoms is only to anchor the pentylamine layer. 
Hence it is not necessary to form a continuous carbon coating of the substrate. 
For example the Pt/C samples produced by D. Klinov were obtained by 
simultaneous evaporation of both Platinum and Carbon in unknown proportions. 
(However in the transport devices for measurement of DNA transport the nature of the 
substrate will not be relevant since the active region will be contaminated by carbon 
from the focused ion beam microscope). 

\section{Combing DNA molecules onto electrodes with the pentylamine technique}
\label{chap:dnacomb}

In the previous section we described how DNA molecules can be attached on a metallic film 
with the pentylamine plasma functionalization. From our experiments it seems that this procedure 
is successful at attaching DNA molecules only on carbon coated substrates. In this section 
the substrate will be a Platinum film (3 nm) sputtered on a cleaved mica surface and covered 
by a layer ($\simeq 10\;{\rm nm}$ of evaporated amorphous carbon).

\begin{figure}[ht]
\centering
\includegraphics[width=0.7 \columnwidth]{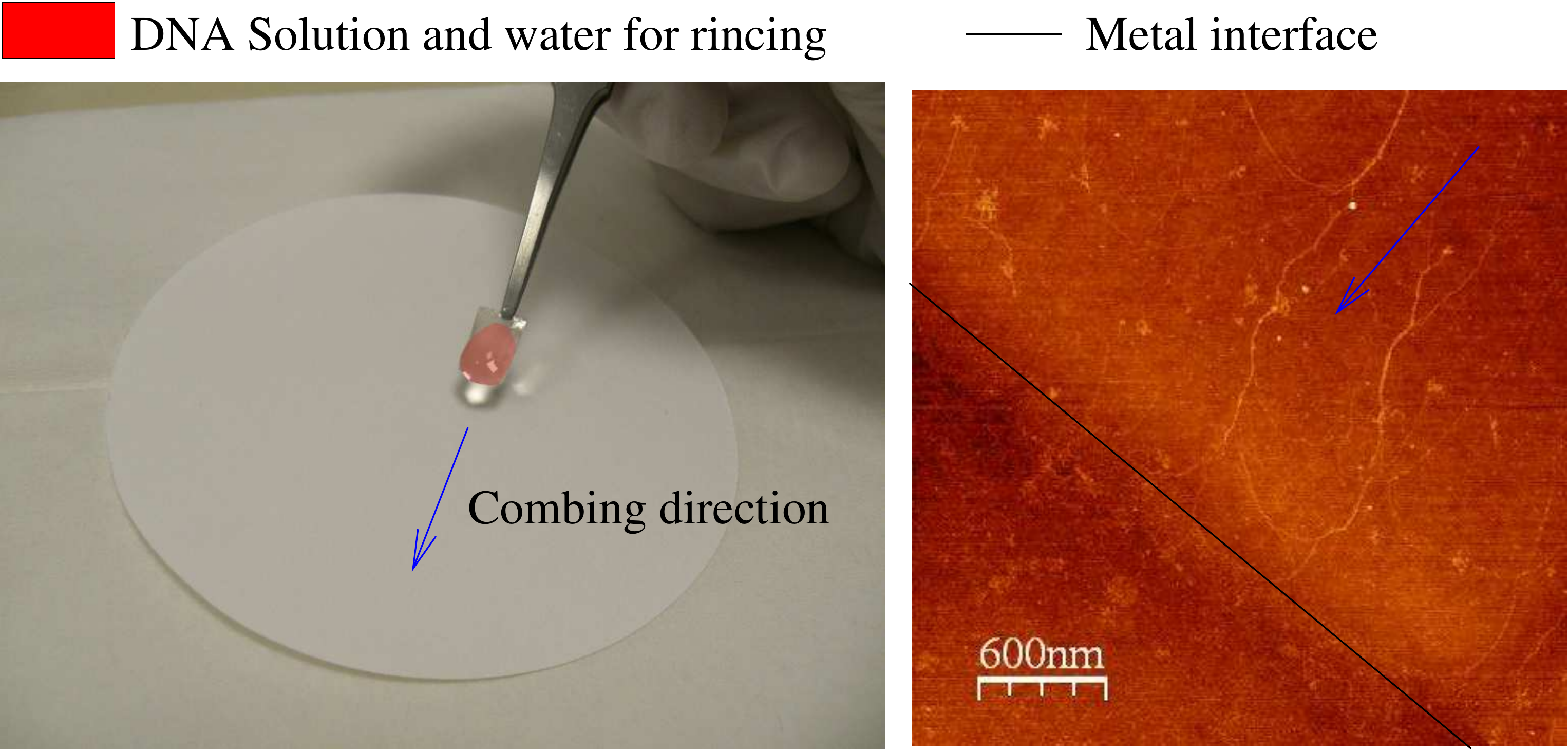}
\caption{Combing of DNA molecules using a drying water air interface. A droplet of DNA ($\simeq 15\;{\mu L}$) is incubated 
on the functionalized surface for a minute. Afterward $\simeq 100\;{\mu L}$ of water are added to the drop to avoid 
formation of salt crystals when the solution is dried. The combing is achieved by draining the liquid from the 
sample with a filter paper (combing direction indicated by blue arrows, see photography on the left). 
An AFM image of the molecules near metal-insulator interface is shown on the right. 
Molecules are combed on the metal electrodes but turn around near the interface, and no molecule 
crosses the interface in this picture. 
}
\label{fig:CombDry}
\end{figure}

We now describe how to orient DNA molecules perpendicularly to an insulating gap separating wide metallic electrodes.
The possibility to orient individual DNA molecules with a moving air-liquid interface was first 
established experimentally in Ref. \cite{DnaCombing}. Figure ~\ref{fig:CombDry} shows a photography 
of the deposition setup, and an AFM image of the molecules at the metal-insulator interface 
(both were covered by an amorphous carbon layer before deposition). Even if molecules are combed 
on the electrodes, they seem to turn around at the metal boundary avoiding the insulator. 
This guiding may be explained by pinning of the liquid air-interface at the border between the two regions.
Moreover many molecules are overstretched which is to be avoided for transport measurements. 
For this reason we have chosen an alternative technique which consists in orienting the
molecules with a flow. A macroscopic Poiseuille flow is not perturbed by defects and interfaces 
on the nanoscale, and regulation of the flow velocity allows in principle to control the elongation 
of the molecules. 

\begin{figure}[ht]
\centering
\includegraphics[width= 0.7 \columnwidth]{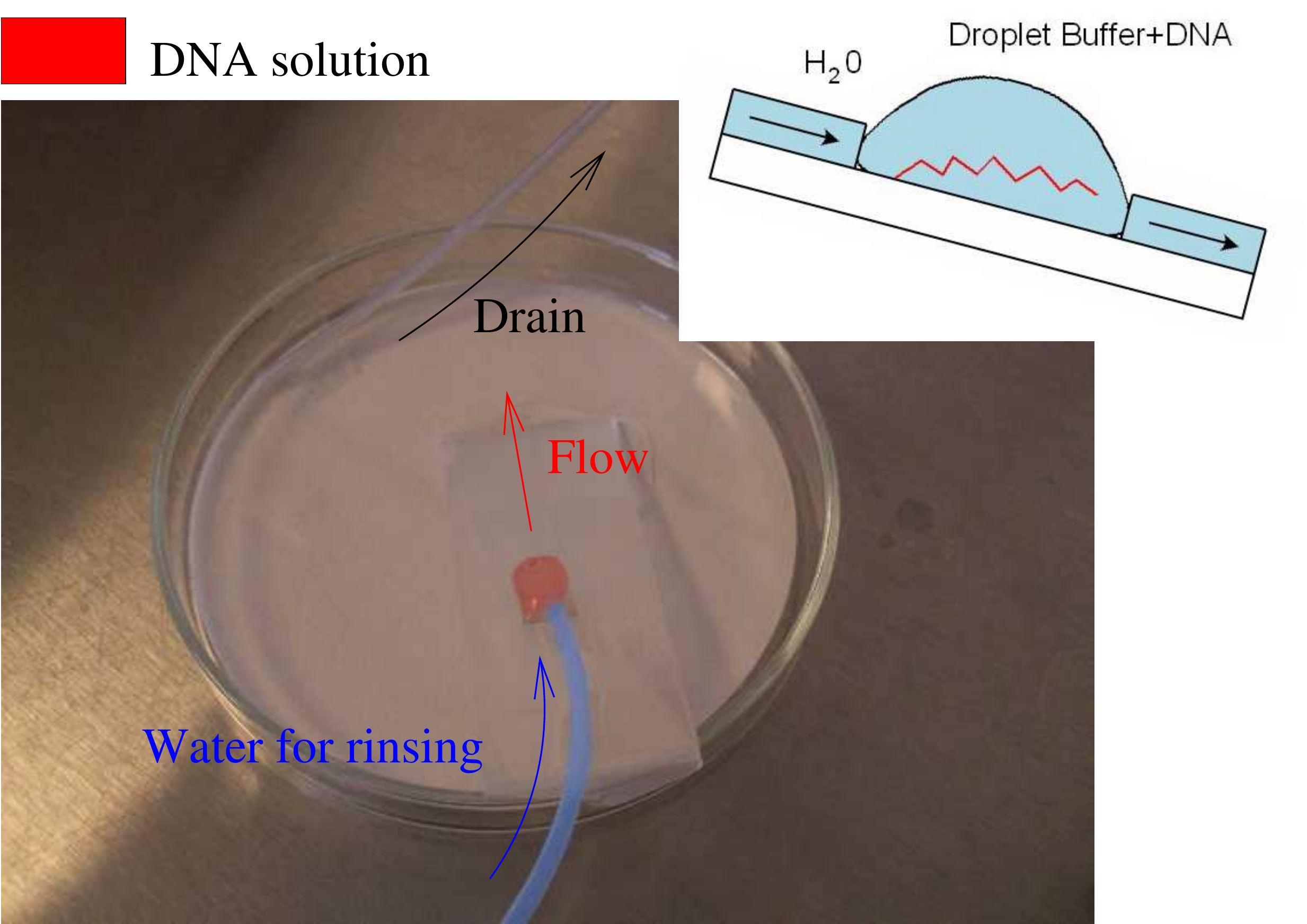}
\caption{Combing of DNA molecules with a flow. A drop of DNA solution is deposited on the sample and incubated for 
a few minutes. The substrate is then rinsed with a flow of pure water injected by a peristaltic pump.
The liquid regularly runs off the sample, which is tilted in the direction of the flow
allowing to collect the excess  liquid in a petri dish under the sample. 
The liquid is drained by the peristaltic pump thus keeping the water level constant. 
}
\label{fig:CombFlow}
\end{figure}

In our deposition protocol (see Fig.~\ref{fig:CombFlow} for a description of the setup) 
the droplet is first incubated on the sample surface for around a minute. 
This allows DNA molecules to bind to the surface at a few 
contact points without complete adsorption on the surface. Afterward the rinsing flow is turned on,
it provides a flow rate of $\simeq 5\;{\rm ml/min}$ of pure water on the sample. 
This flow has a double function of combing DNA molecules and rinsing salt residues that 
may form on the surface rendering AFM imaging difficult. 
It is interesting to know in which flow speed regime DNA molecules may start to be overstretched. 
A quantitative study of the dynamics of a tethered DNA molecules under a Poiseuille flow 
was performed recently \cite{DNAflowENS} using fluorescence microscopy. 

It was found that the elongation of the molecules was governed by a single dimensionless parameter,
called the Weissenberg number ${\rm Wi} = {\dot \gamma} \tau$ where ${\dot \gamma}$ is the shear rate 
and $\tau$ is the longest relaxation time of the DNA molecule (it was found to be $\tau \simeq 0.4\;{\rm s}$ 
for  $\lambda$-DNA labeled  with fluorescent beads \cite{DNAflowENS}). Complete elongation of the 
molecules occurred only for ${\rm Wi} \simeq 100$, while an elongation of $25 \%$ occurs already for 
${\rm Wi} \simeq 5$. The shear rate in our experiments can be estimated as follows: the outflow of liquid 
on the surface is $D_{flow} \simeq 5\;{\rm ml/min}$, for a cross section of the droplet 
of the order of $S = H \times (2R)$ where $H = 1.5\;{\rm mm}$ is the droplet height 
and $R = 5\;{\rm mm}$ is the droplet radius. 
The mean velocity in the fluid is $V = D / S$ and since the flow vanishes at the contact with the substrate
the shear rate is ${\dot \gamma} = V / H \simeq 4\;{\rm s^{-1}}$ leading to ${\rm Wi} = 1.5$. 
This calculation shows that with our typical flow parameters we are far from the threshold ${\rm Wi} \simeq 100$ 
where molecules may start to be overstretched. As a result we have set the debit to a value around $5\;{\rm ml/min}$ 
where the flow on the sample was stable without risks of uncontrolled drying of the drop during the rinsing process.

\begin{figure}[ht]
\centering
\includegraphics[width= 0.7 \columnwidth]{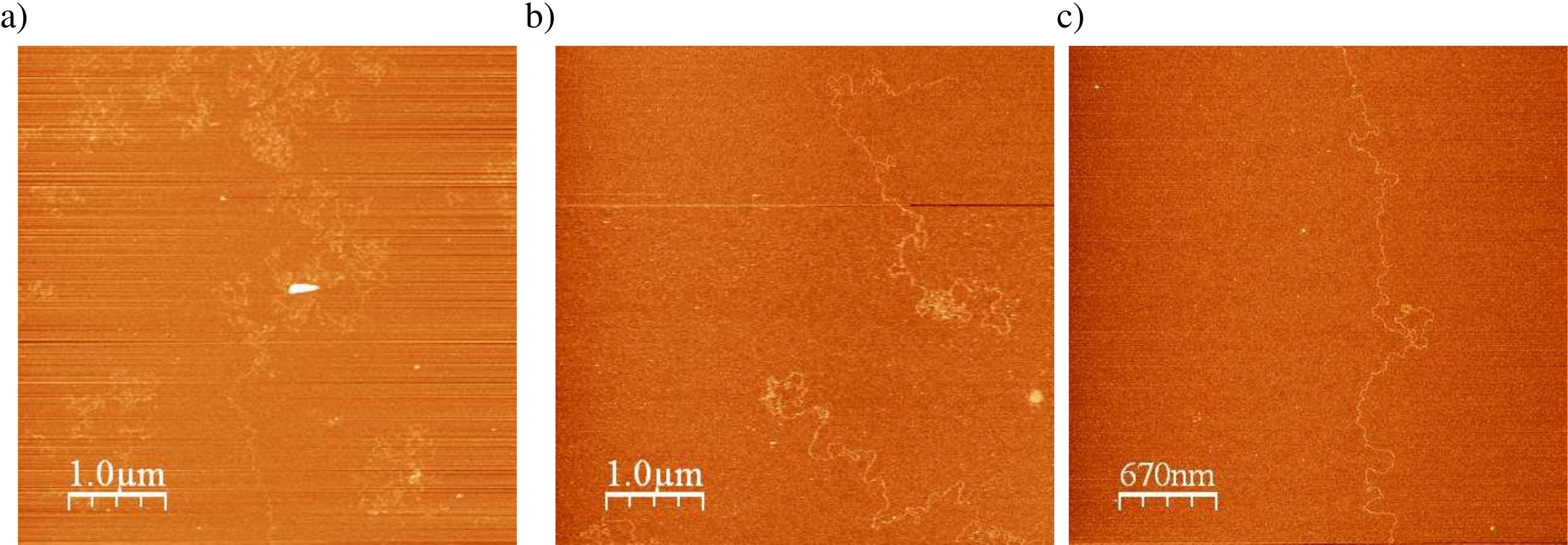}
\caption{AFM images of DNA molecules combed on a platinum carbon pentylamine substrate with a liquid flow.
Images from left to right correspond to increasing flow velocity.
}
\label{fig:CombFlowPent}
\end{figure}

AFM images of DNA molecules deposited with this technique on a carbon coated platinum surface are 
shown on Fig.~\ref{fig:CombFlowPent}. The extension of molecule 
depends on the number of active binding sites created during the glow discharge, 
the incubation time before rinsing and on the water flow rate. Since all these parameters 
are hard to fix in a reproducible way, sometimes molecules are more collapsed 
onto themselves (Fig.~\ref{fig:CombFlowPent}.a) and sometimes they are more extended (Fig.~\ref{fig:CombFlowPent}.c).
However binding and orientation of the molecule  (in the 
range suggested by the different cases of Fig.~\ref{fig:CombFlowPent}) was reproducible 
with this technique. 

\begin{figure}[ht]
\centering
\includegraphics[width= 0.5 \columnwidth]{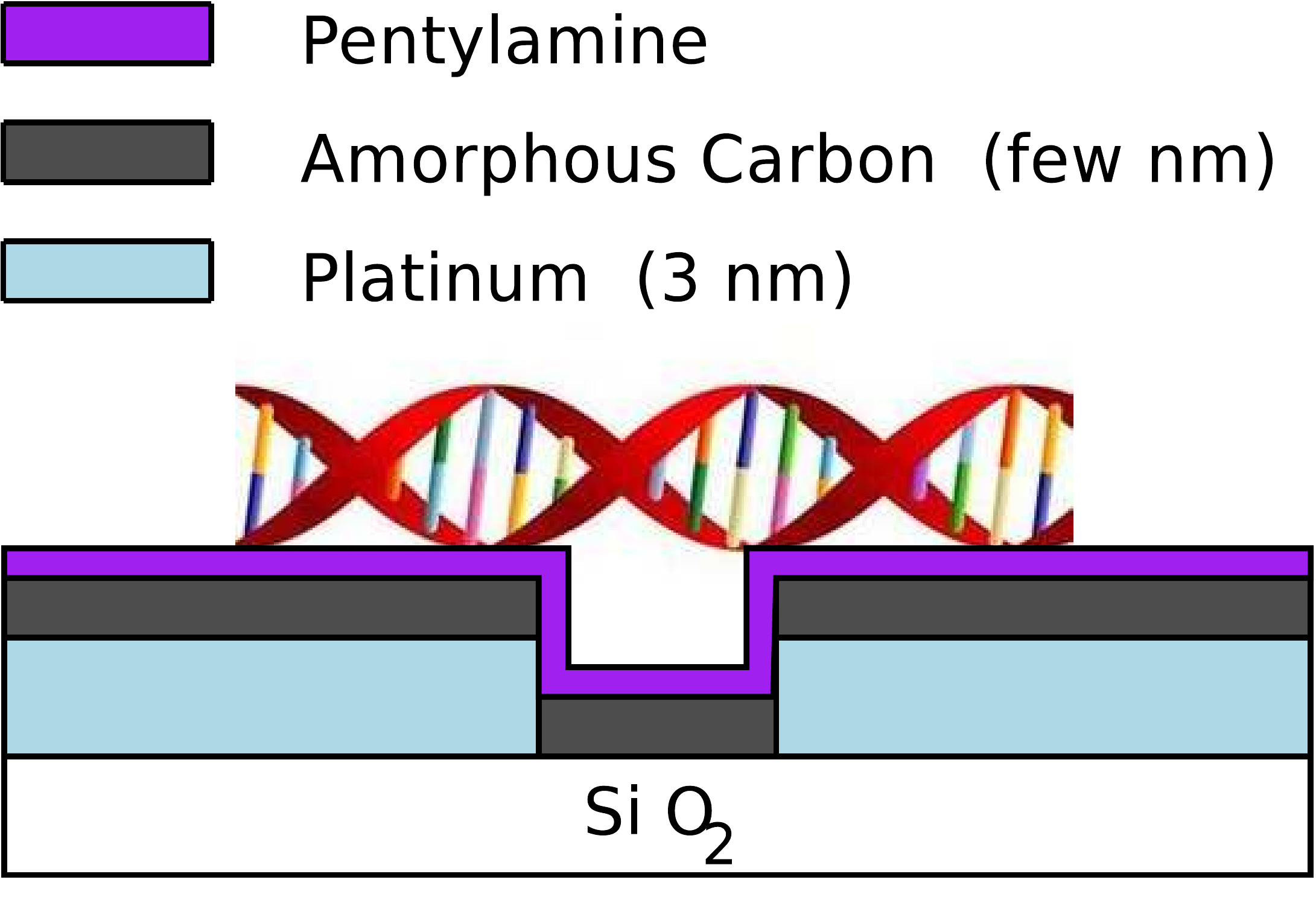}
\caption{Schematic representation of the material layers for samples produced by 
electron beam lithography. Structure of the samples produced with the focused ion beam 
is similar except that the substrate is mica and will be discussed in more detail in Section~\ref{chap:FIB}.
}
\label{fig:LithoADN}
\end{figure}

\begin{figure}[ht]
\centering
\includegraphics[width= 0.7 \columnwidth]{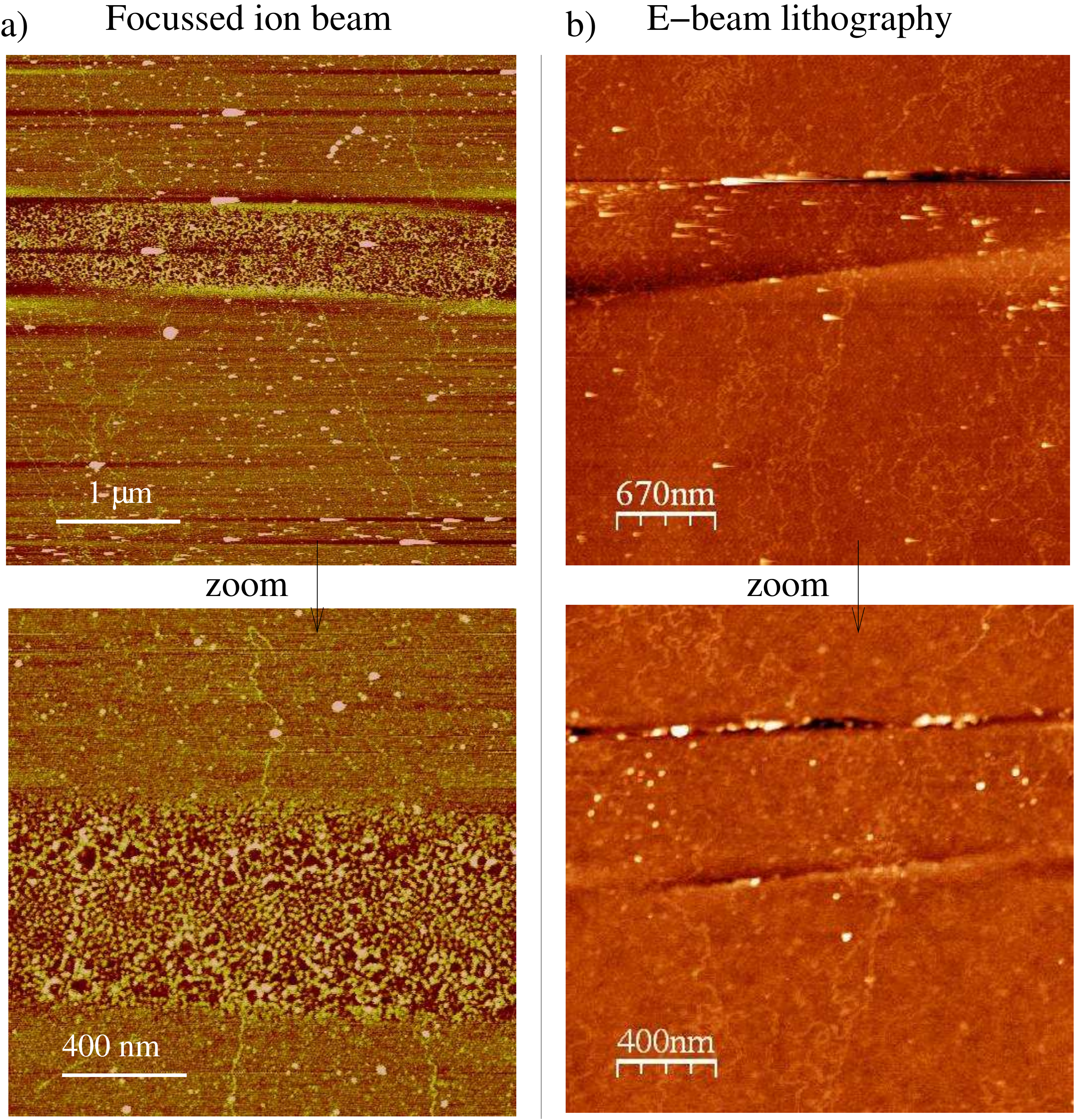}
\caption{AFM image of DNA molecules deposited across insulating gaps using the flow combing technique. 
a) Shows a sample produced with focused ion beam etching. The insulating gap is actually around $100\;{\rm nm}$ wide,
even if apparent width is around 400 nm.
This occurs because the roughness of the Pt film is increased in a large area around the gap after FIB irradiation.
 b) Shows DNA molecules across a gap 
produced with electron beam lithography. Note that the gap realized with electron beam lithography 
is cleaner and one can follow the molecules inside the gap.  This is not possible for the 
gap produced by focused ion beam etching due to the roughness of the substrate after etching 
(see also Section \ref{chap:FIB}).
}
\label{fig:Gap}
\end{figure}

Now all ingredients are assembled to deposit DNA molecules across an insulating gap 
using the pentylamine technique. We have done this with electrodes produced with 
both electron beam lithography and focused ion beam etching (this technique was also used in 
\cite{Klinov} and will be described further on in Section \ref{chap:FIB}).
A cross section of the material layers constituting electrodes and the gap is sketched
on Fig.~\ref{fig:LithoADN}, this structure is similar for both electron-and ion beam processed samples.
The width of the gaps for these samples was between $100$ and  $400\;{\rm nm}$. 
Molecules crossing the gap are clearly visible in Fig.~\ref{fig:Gap} however all the gaps 
where we deposited molecules always remained insulating with resistances larger than Giga-Ohms 
despite presence of carbon and pentylamine layers.
In order to verify that the surface of the electrodes is not insulating due 
to the formation of an oxide layer or due to organic contamination we have deposited 
HIPCO single wall carbon nanotubes (SWNT) from a dichloroethane solution across the electrodes.
The resistance of the junction then dropped to values of the order of $100\;{\rm kOhms}$
suggesting that our electrodes were clean enough to make contacts to nanonotubes.
However we will argue in the next sections that the electrodes were probably covered 
by an insulating pentylamine layer during deposition of DNA molecules.
The organic solvent of the SWNT may have cleaned the electrode surface 
thereby allowing the formation of an electrical contact.

These results are in disagreement with findings from \cite{Kasumov,Klinov} which suggested that 
the presence of a pentylamine layer creates a suitable substrate where conduction of DNA molecules 
is possible. However the above experiments depend on many parameters which are often poorly defined. 
In the absence of any credible indication on the origin of this discrepancy 
we have asked D. Klinov to deposit DNA molecules with his setup on samples 
with insulating gaps prepared in our laboratory using a focused ion beam
reproducing as accurately as possible the experiment from Ref. \cite{Klinov}.

\clearpage 

\section{Fabrication of narrow insulating gaps using a focused ion beam}
\label{chap:FIB}

\begin{figure}[ht]
\centering
\includegraphics[width= 0.7 \columnwidth]{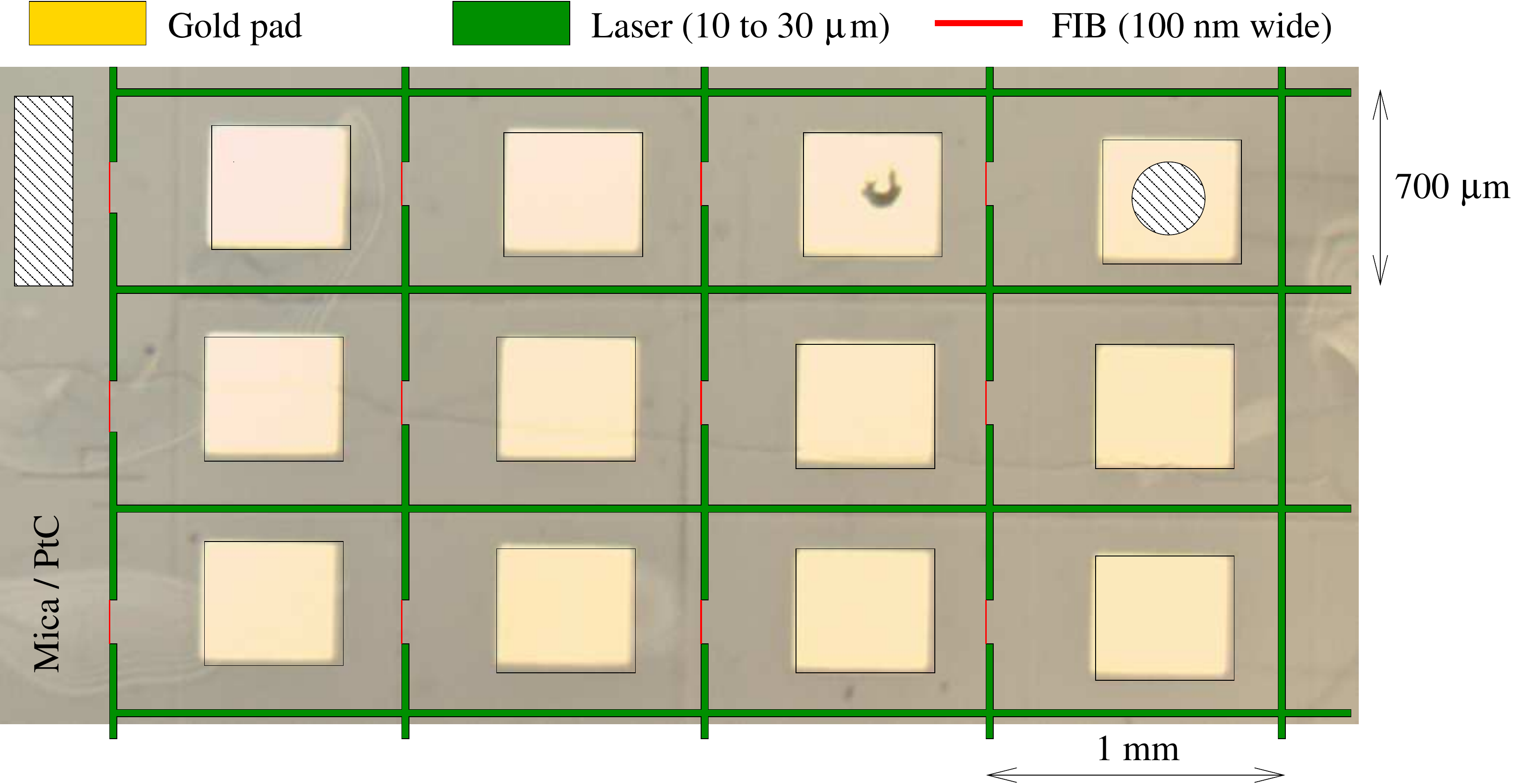}
\caption{Sketch of a sample produced with FIB etching with a sample photography on the background.
It shows the gold contact pads, the pattern exposed to laser irradiation (green line)
and the small gaps etched with FIB at the last step (red line). The scales are not all preserved 
on this diagram: the red line (region etched with FIB) is in reality $50\;{\rm \mu m}$ long 
and $100\;{\rm nm}$  whereas the width of the lines etched with laser is roughly $10\;{\rm \mu m}$.
}
\label{fig:Layout}
\end{figure}

This method of fabrication of insulating gaps does not use 
electron beam lithography and therefore avoids the contamination 
of the surface by residues from organic resist. 
A thin layer of platinum carbon was deposited in D. Klinov's laboratory on a freshly cleaved mica substrate.
This metal layer has an estimated thickness of at around $5\;{\rm nm}$ and a resistance per square of 
around $1\;{\rm k\Omega}$. 
A schematic representation of the sample layout after laser and focused ion beam (FIB) etching 
is shown on Fig.~\ref{fig:Layout}.
In a first step thick gold ($\simeq 200\;{\rm nm}$) 
contact pads were evaporated through a mechanical mask. 
We then cut long trenches in the metallic film using an ultraviolet focused laser 
with spot-size around $10$ to $30\;{\rm \mu m}$. 
The laser locally heats the surface and evaporates the metal layer creating holes 
in the metal of the order of the spot size. Programmable motors then allow to 
expose the metal in predefined patterns around the golden contacts 
leaving regularly spaced metal openings $60\;{\mu m}$ long. 
These remaining metal stripes were opened with a Gallium FIB which can etch narrow 
 $100\;{\rm nm}$  wide trenches. In order to determine the minimal dose of FIB irradiation
required to produce a narrow insulating gap in the platinum film we have developed a technique 
for in-situ measurement of the film resistance inside the FIB microscope. 

\begin{figure}[ht]
\centering
\includegraphics[width= 0.7 \columnwidth]{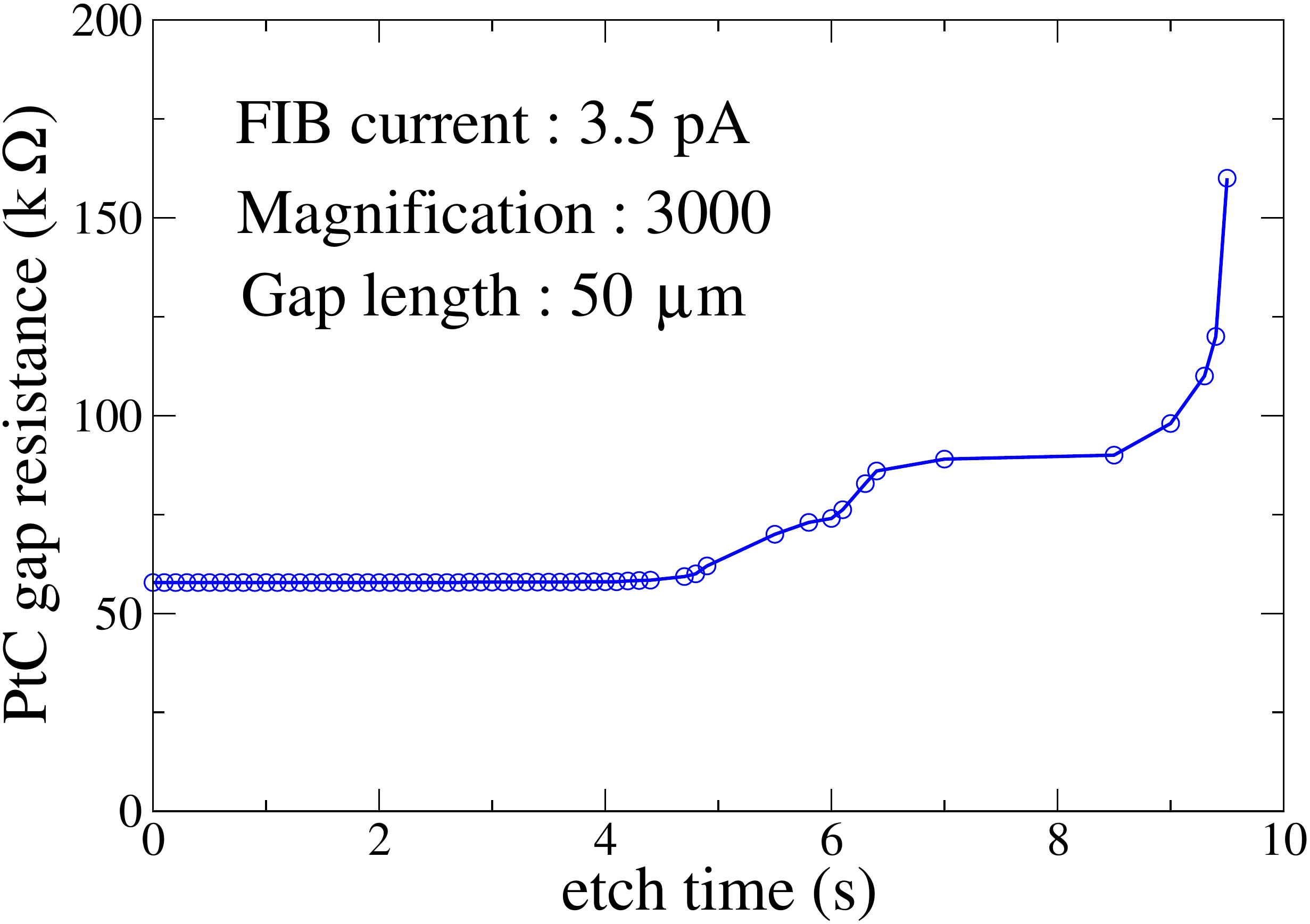}
\caption{Resistance of a gap during FIB etching as a function of exposure time. The gap is etched in a single scan mode 
with a scan time of $0.1\;{\rm s}$ which allows to measure the resistance after each scan. After total time $t > 9.7\;{\rm s}$
the resistance jumps and the gap becomes insulating. Usually at the beginning of the etching there is a short 
phase were conduction drops by a small amount that can not be seen on this scale. 
This surprising behavior will be discussed in Section \ref{chap:FibAFMSEM}.
}
\label{fig:FibEtchTime}
\end{figure}

This technique allows us to continuously monitor the resistance between several gaps 
contacted in series (for e.g. between the hatched regions on Fig.~\ref{fig:Layout}).
First an estimate of the required etch time is obtained by cutting half of one gap and 
monitoring simultaneously the increase of resistance as a function of time. 
We estimate the optimum etching time of the PtC film from the value at which the resistance saturates. 
After this operation the measured resistance is still finite since 
only half of the gap is etched, this allows to repeat the operation and accumulate statistics on a few (typically 5) gaps. 
At the last step one of the gap is opened completely until resistance diverges, the resistance dependence on the etch time 
is shown on Fig.~\ref{fig:FibEtchTime} for one of the junctions. This last measure gives a very precise estimate 
of the minimal dose. Considering possible fluctuations in the thickness of the film, we increase this dose by around 
20\% and etch all the remaining gaps  with the same dose. 

\begin{figure}[ht]
\centering
\includegraphics[width=0.6 \columnwidth]{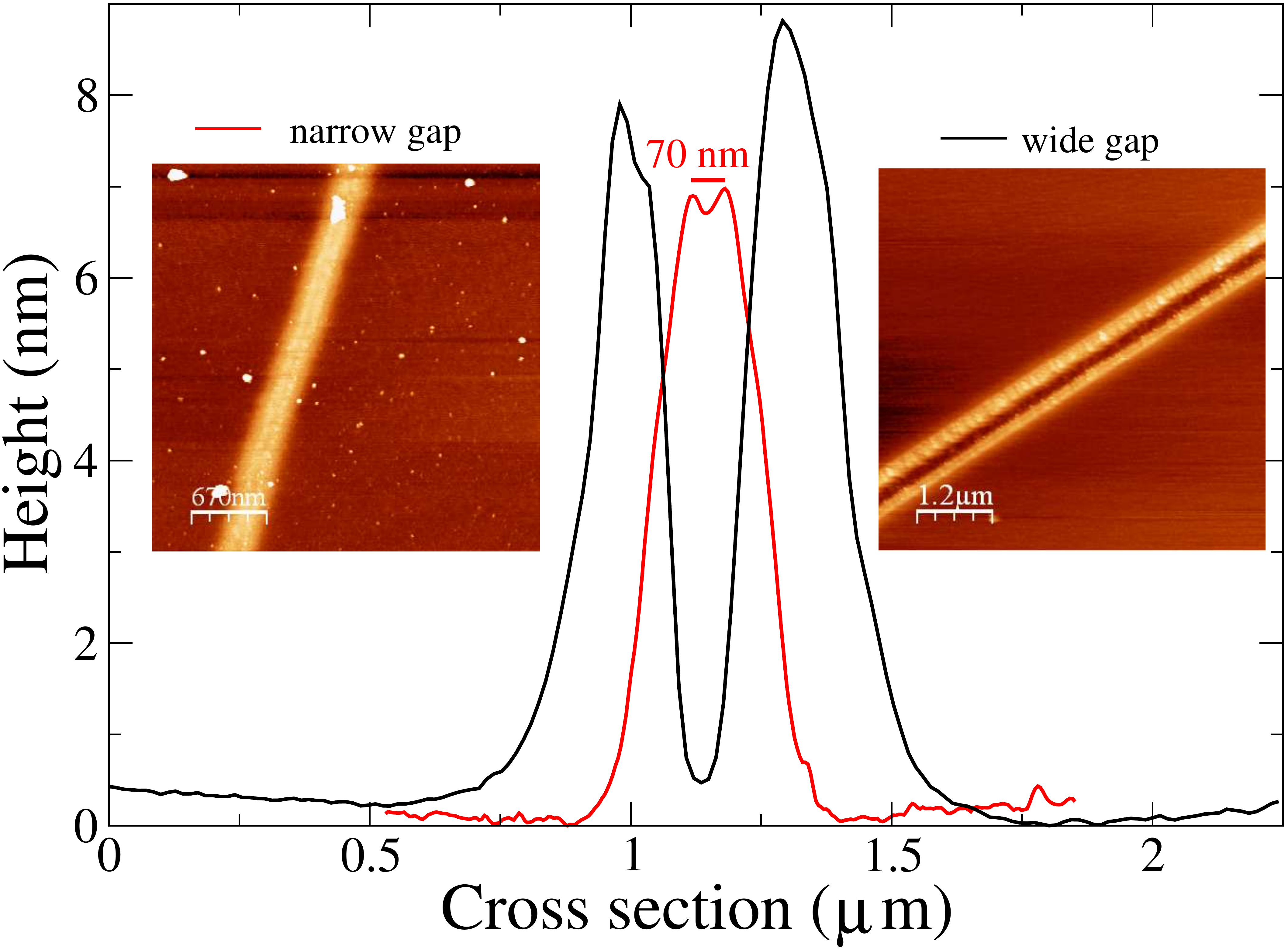}
\caption{
Averaged height profile from two gaps prepared using FIB, their AFM images are shown in the color insets. 
The narrow gap was obtained during the calibration resistance measurement from 
Fig.~\ref{fig:FibEtchTime} while the wide gap was obtained with a larger exposure time. 
It is difficult to distinguish the insulating region from the main contrast that comes from the 
roughness near the edge of the gap produced during the FIB etching.
}
\label{fig:FibAFM}
\end{figure}

Atomic force microscopy images of the gaps fabricated with the FIB are shown on Fig.~\ref{fig:FibAFM}. 
For the narrowest gap the height profile exhibits a characteristic peak with a small dip 
in the center where the gap is insulating. Certainly the dip is actually deeper than shown
on the AFM image because of tip convolution effects.
At higher irradiation doses, the gap grows in size and more pronounced side peaks appear. 

Surprisingly for the minimal etch time, the width of the insulating region is of the order of $\simeq 70\;{\rm nm}$ 
and is very slender compared to the contamination peak which extends over about $500\;{\rm nm}$. 
The cross section of the peak for the narrow gap is $\Sigma = (250\;{\rm nm} - 70\;{\rm nm}) \times 7\;{\rm nm} \simeq 1260\;{\rm nm^2}$  
where $7\;{\rm nm}$ is the gap height, $250\;{\rm nm}$ and $70\;{\rm nm}$  are respectively the bump and gap widths.
Assuming that the thickness of the platinum film is around $5\;{\rm nm}$ we find that the cross section of the excavated platinum is sensibly 
smaller $70\;{\rm nm} \times 5 \; 350 {\rm nm^2}$. It is possible that FIB dug also into mica 
providing the remaining material inside the bump. However since we stopped etching just when the gap became insulating,
it is unlikely that we dug very deep into the mica surface. 
It is instructive to compare the number of atoms ${\cal N}$ inside this ``bump'' 
(we find that ${\cal N} = \Sigma \times 50 \; {\rm \mu m} / (0.25 \;{\rm nm})^3 \simeq 4 \;\times\; 10^{9}$ 
where $50 \; {\rm \mu m}$ is the gap length and $0.25\;{\rm nm}$ is the average distance between atoms for e.g. in a gallium crystal),
with the number of gallium atoms ${\cal N}_{FIB}$ sent by the FIB. 
This number can be estimated from the FIB current and the etch time for the narrow gap on Fig.~\ref{fig:FibEtchTime},
it amounts to ${\cal N}_{FIB} = 3.5\;{\rm pA} \times 10\;{\rm s} / e \simeq 2 \; \times 10^{8}$ (e is the elementary charge). 
According to this calculation there is an almost two order of magnitude difference 
between the quantity of atoms in the bump ${\cal N}$ and the quantity of atoms emitted by the ion beam ${\cal N}_{FIB}$.
Although these estimates are not very accurate they suggest that there may be a third origin for the peaks around the gap
other than displacement of platinum and mica  and injection of gallium atoms. 
It is possible that the metallic film dewets locally from the mica surface due to heating by FIB 
creating a bump (we have remarked adhesion of thin films is generally not very good on mica). 
In all cases AFM scans in the direction parallel to the gap show ( see AFM images from Fig.~\ref{fig:Gap}.a )
that the surface is very rough in a region extending across $\simeq 700\;{\rm nm}$ much wider than the insulating region ($\simeq 200\;{\rm nm}$),
indicating that the PtC film is damaged or contaminated in a large region outside the gap. 

For the moment we stop here the analysis of the gap structure (but it continues in Section \ref{chap:FibAFMSEM} !). 
In the discussion that follows, it is enough to keep in mind that the structure of the gap 
may actually be quite complicated even if it is obtained by ``physical''  means without organic resists. 

A total of about twelve substrates were sent to D. Klinov in Moscow for the deposition of DNA molecules,
with about ten gaps opened by FIB on each sample.
The deposition protocol used by D. Klinov seems similar to the protocol we used in our
deposition experiments (see Section~\ref{chap:dnacomb}). The only difference is
that in D. Klinov's pentylamine setup a gas flow continuously
refreshes the pentylamine gas in the discharge chamber. This procedure can reduce the probability 
of forming radicals/ions which require breaking several bonds or removing several electrons from the pentylamine molecule.
But this probability is low anyway and it is not clear how it could influence the conduction of DNA. 

The results on conduction after deposition of DNA molecules in D. Klinov's laboratory are summarized in Table~\ref{tab:Samples}.
On three samples conduction was observed after deposition of DNA molecules across slits that were 
insulating before DNA deposition. The statistics on these three sample is very favorable since 
11 out of 15 slits became conducting. Moreover the deposition protocol was repeated with a buffer solution 
without DNA on a control sample from the same batch as two of the conducting samples 
(the mica sample was cut in three pieces before the pentylamine treatment) and all the 14 gaps remained insulating.
This statistic is strongly in favor of an interpretation in term of conduction through DNA molecules.
However this statistical argument must be handled with care. If we consider all the samples were deposition 
of $\lambda$ molecules was attempted, the conducting slits represent only around 10\% of the prepared structures.
On the other samples or no DNA molecules  could be detected with AFM indicating that pentylamine 
functionalization was not effective or the molecules were insulating.  Hence  observation of conductivity after 
attempts of DNA deposition has actually a low success rate even when deposition is done by D. Klinov following 
the recipes used in \cite{Kasumov,Klinov}. We have reached the conclusion that one of the reasons behind this irreducibility 
lies in the structure of the sample after etching which will be discussed in more detail in Section \ref{chap:FibAFMSEM}.
We now turn to transport measurements on the three samples where conduction was observed.

\begin{table}
        \begin{center}
                \begin{tabular}{|c|c|}
\hline
		  Number of substrates & 12  \\
\hline
		  Number of FIB  slits & $\simeq$ 100 \\
\hline
                  Number of substrates with visible $\lambda$ DNA   & 5  \\
\hline
		  Number of substrates with conducting slits after $\lambda$ deposition  & 3  \\
\hline
		  Number of conducting slits after $\lambda$ deposition & 11  \\
\hline
		  Number of slits on these three subtrates & 15    \\
\hline
                  Number of slits on the control sample & 14 \\
\hline
		  Number of conducting slits after buffer & 0 \\
\hline   
		\end{tabular}
        \caption{Success rates for the formation of conductive junctions by deposition of $\lambda$ molecules. }
        \label{tab:Samples}
        \end{center}
\end{table}

\section{Transport measurements on conducting DNA samples}
\label{chap:DnaTransport}

\begin{figure}[ht]
\centering
\includegraphics[width=0.7 \columnwidth]{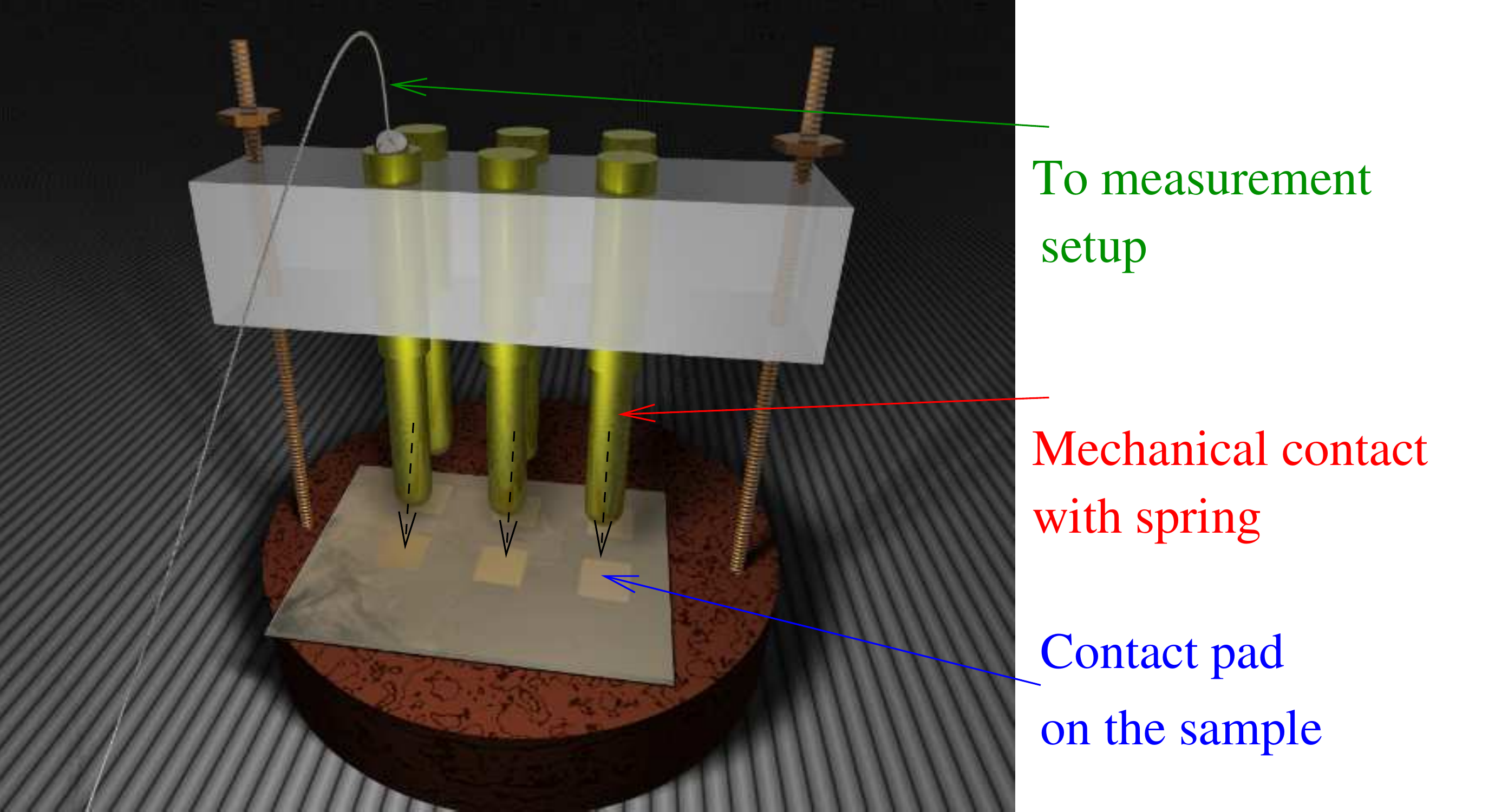}
\caption{
Three dimensional model of the mechanical connection system we used to contact our DNA sample 
to the dilution fridge.
}
\label{fig:FigPogo}
\end{figure}

Before performing transport measurements on the three substrates where conduction appeared 
after deposition of DNA molecules we had to connect the samples to a sample holder 
which can be mounted inside one of our dilution fridges. This connection can be realized 
through thin ($\simeq 20\;{\mu m}$ diameter) wires with ultrasound bonding 
or glued with silver paint. Ultrasound bonding on a sample with three conducting gaps,
led to disappearance of conduction on three gaps. It is possible that an electrical discharge 
was created during the ultrasound  bonding destroying the conduction across our samples. 
In order to avoid this discharge we have decided to contact the second sample using silver paint. 
Surprisingly with silver paint conduction was also destroyed on the five conducting gaps of the second sample. 
During the process of contacting the gaps we checked their conductivity under a test-probe several times. 
The resistance of a gap could change even when the silver paint drops were deposited on the 
contact pads of the other samples. For example the resistance across one of the gaps took the following values: 
$1.8\;{\rm k \Omega} \rightarrow 160\;{\rm k \Omega} \rightarrow > 1000\;{\rm k \Omega} \rightarrow 5\;{\rm k \Omega} \rightarrow \infty$. 
These observations suggested that our samples were sensible to the vapors of the silver paint solvent.
Hence we decided to avoid silver paint for the contacts on the last sample. 
A possibility was to replace silver paint with indium paste however the latter did not stick to 
contact pads after the pentylamine discharge probably because of the presence of the organic layer. 
A. Kasumov then proposed to use a system of mechanical contacts with springs that we 
fabricated specially to fit the geometric parameters of the last remaining sample (see Fig.~\ref{fig:FigPogo}).

This allowed us to contact 5 conducting gaps which we measured at low temperature.
Their resistance at room temperature is listed in Tab.~\ref{tab:DNAtrans}

\begin{table}
        \begin{center}
                \begin{tabular}{|c|ccccc|}
\hline
	measured at low temperature & $ 3\;{\rm k \Omega} $ & $ 4.8\;{\rm k \Omega} $ & $ 6.1\;{\rm k \Omega} $ & $ 10\;{\rm k \Omega} $ & $ 10\;{\rm M \Omega} $ \\ 
\hline
        measured only at room temperature &  $ 4.0\;{\rm k \Omega} $ & $ 8.2\;{\rm k \Omega} $ & $\infty$   &  $\infty$  &  \\
\hline
		\end{tabular}
        \caption{Room temperature resistances of the gaps at room temperature after deposition of DNA molecules on the sample measured at low temperature. }
        \label{tab:DNAtrans}
        \end{center}
\end{table}

\begin{figure}[ht]
\centering
\includegraphics*[width=0.6 \columnwidth]{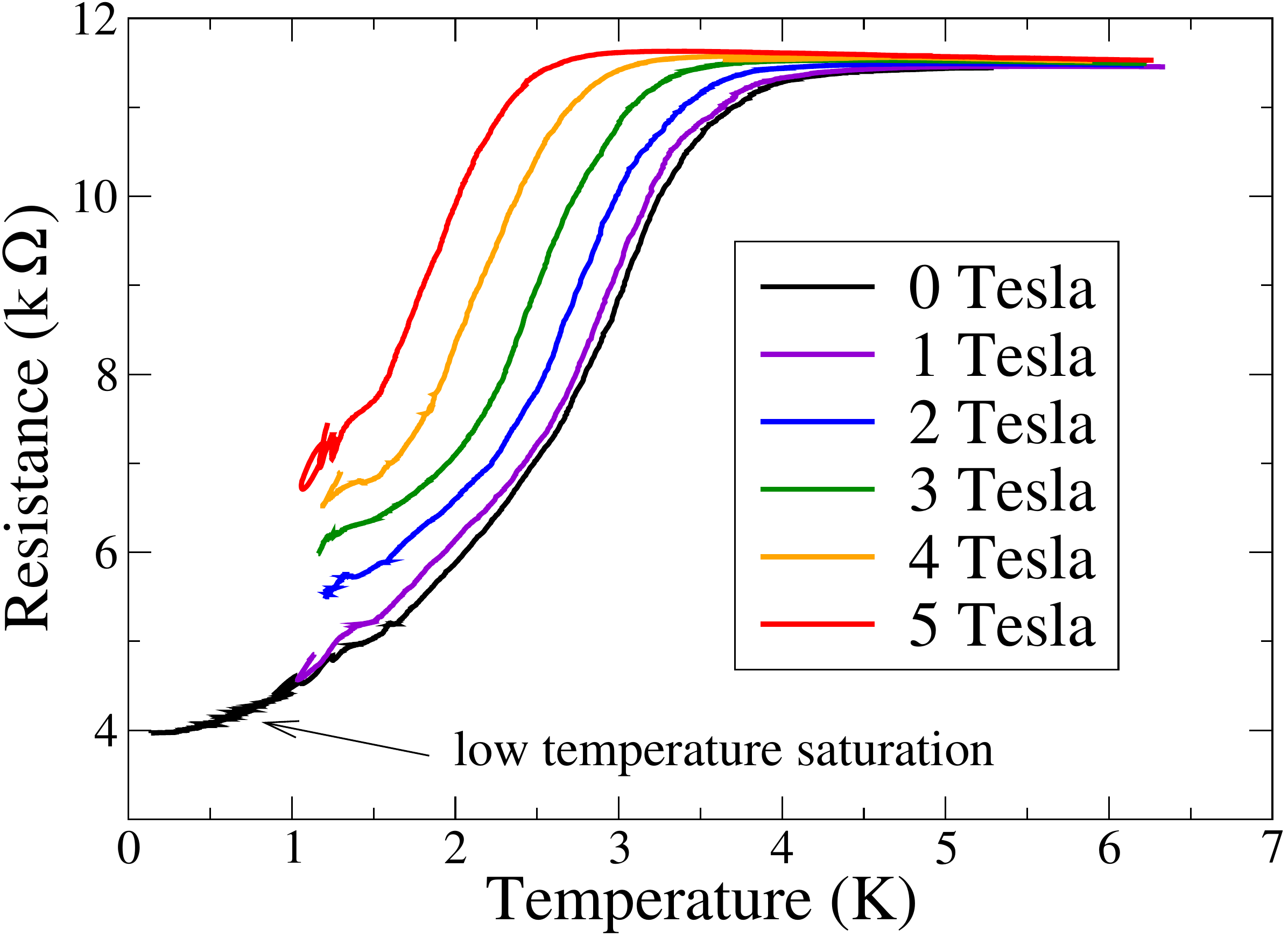}
\caption{
Temperature dependence of resistance for the $10\;{\rm k \Omega}$ junction (see Tab.~\ref{tab:DNAtrans}) at several magnetic fields. 
}
\label{fig:FigDnaTemp}
\end{figure}

\begin{figure}[ht]
\centering
\includegraphics*[width=0.6 \columnwidth]{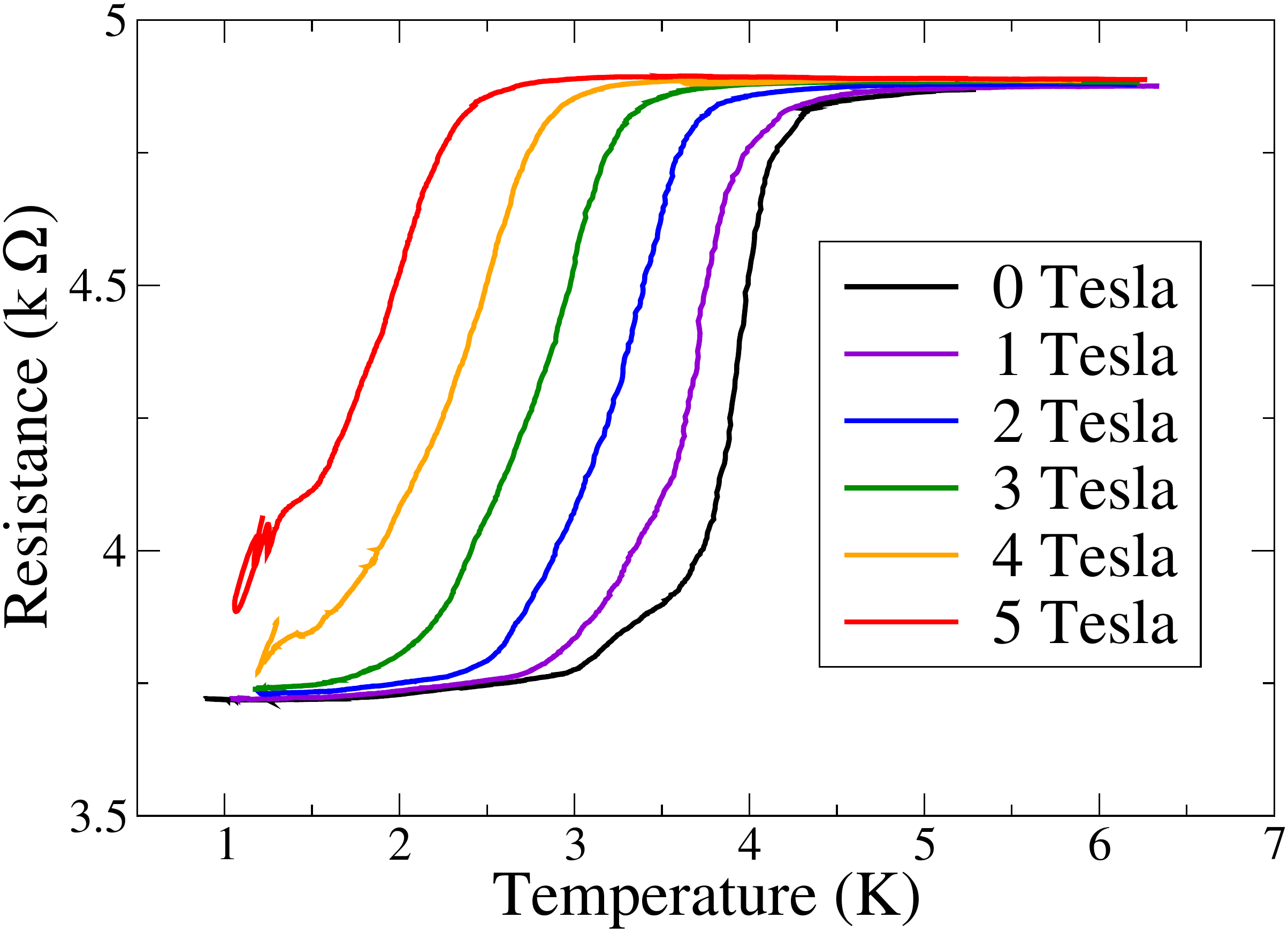}
\caption{
Temperature dependence of resistance for the $4.8\;{\rm k \Omega}$ junction (see Tab.~\ref{tab:DNAtrans}) at several magnetic fields. 
}
\label{fig:FigDnaTemp4Kohms}
\end{figure}

To our surprise all four samples with room temperature resistance $\le 10\;{\rm k \Omega}$ exhibited superconducting behavior
at low temperature. Indeed contrarily to the experiment \cite{Kasumov} where superconducting electrodes where 
used to contact the DNA molecules, our platinum/carbon contacts are in a normal state. 
The dependence on temperature for different magnetic fields is shown for two samples on 
Figs.~(\ref{fig:FigDnaTemp},\ref{fig:FigDnaTemp4Kohms}). Superconductivity appeared below $4\;{\rm K}$ 
and results in a drop of resistance that saturates at low temperature because of the finite 
resistance of the normal contacts. The 10 ${\rm k \Omega}$ sample displayed a very smooth transition 
as a function of temperature (see Fig.~\ref{fig:FigDnaTemp}) and the drop of resistance saturated at 
$T \simeq 200\;{\rm mK}$. In contrast less resistive samples had a sharper transition (see for e.g. Fig.~\ref{fig:FigDnaTemp4Kohms})
with a saturation temperature around $2\;{\rm K}$. The smooth transition observed in the 10 ${\rm k \Omega}$ sample 
has some similarities to smooth transitions observed in Superconductor-Normal-Superconductor (SNS) junctions in the intermediate 
regime between a long and a short junction. A long SNS junctions is characterized by the presence of 
two transitions. The transition at the highest temperature, stems from the transition of the superconducting contacts 
while at a lower temperature proximity induced superconductivity sets-in in the normal region \cite{Lionel}. 
When the length of the normal part is decreased the two transitions merge into a single smoother transition 
\cite{KasumovSWNT,KasumovSWNT2}. Since in our samples only a single transition is observed, this suggests 
that we have created an SNS junction in this intermediate regime. 

A possible origin of the superconductivity is the Gallium contamination deposited by the FIB.
An insight on the size of the superconducting contamination islands is provided by the dependence 
on magnetic field. Under a magnetic field the resistance of the samples increased, however the maximal  magnetic field available in our 
setup ($5\;{\rm Tesla}$) was not sufficient to completely destroy superconductivity at low temperature indicating a
critical magnetic field of the order of $10\;{\rm Tesla}$ (see Figs. ~\ref{fig:FigDnaTemp} and \ref{fig:FigDnaTemp4Kohms}). 
Assuming the presence of superconducting nanoparticles we can also understand the origin of this relatively high critical field. 
The  magnetic field destroys  superconductivity in a nanoparticle 
when it creates a flux of the order of the flux quantum through the nanoparticle surface $\pi R^2$ where $R$ is the nanoparticle radius.
This criterion gives a typical nanoparticle radius of the order of $R \simeq 10\;{\rm nm}$. 
Through extensive AFM/SEM characterization of the measured sample (see Section \ref{chap:FibAFMSEM})
we were able to establish that these nanoparticles were deposited by FIB. 
Gallium is superconducting with transition temperature $\simeq 1\;{\rm K}$, the presence of carbon impurities 
may increase this transition temperature to $\simeq 4\;{\rm K}$. Indeed tungsten wires deposited with FIB 
containing Gallium and carbon have critical temperature $T \simeq 4\;{\rm K}$ whereas the pure tungsten 
has critical temperature around $< 50\;{\rm mK}$ \cite{Kasumov2}. Thus it is probable that 
the source of superconductivity are superconducting nanoparticles inside (and even outside !) the gap cut by the FIB. 
The nanoparticles themselves can not give rise to ohmic resistances of a few ${\rm kOhms}$,
hence the large amplitude of the resistance (drop for e.g. from $10\; {\rm k\Omega}$ to $3\; {\rm k\Omega}$) between 
the normal and the superconducting states, indicates a configuration where a normal nanowire connecting 
the contacts is rendered superconducting by proximity effect from the nanoparticles.  
To conclude on the dependence on magnetic field, one of the samples displayed 
SQUID like modulation in the magnetoresistance with a period of $0.5\;{\rm Tesla}$ at temperature $T \simeq 2\;{\rm K}$.
These oscillations disappeared at lower and higher temperatures  $T < 1\;{\rm K}$ and $T > 3\;{\rm K}$ suggesting 
a complex geometry with several SNS junctions connected in series and/or in parallel. 

\begin{figure}[ht]
\centering
\includegraphics[width=0.7 \columnwidth]{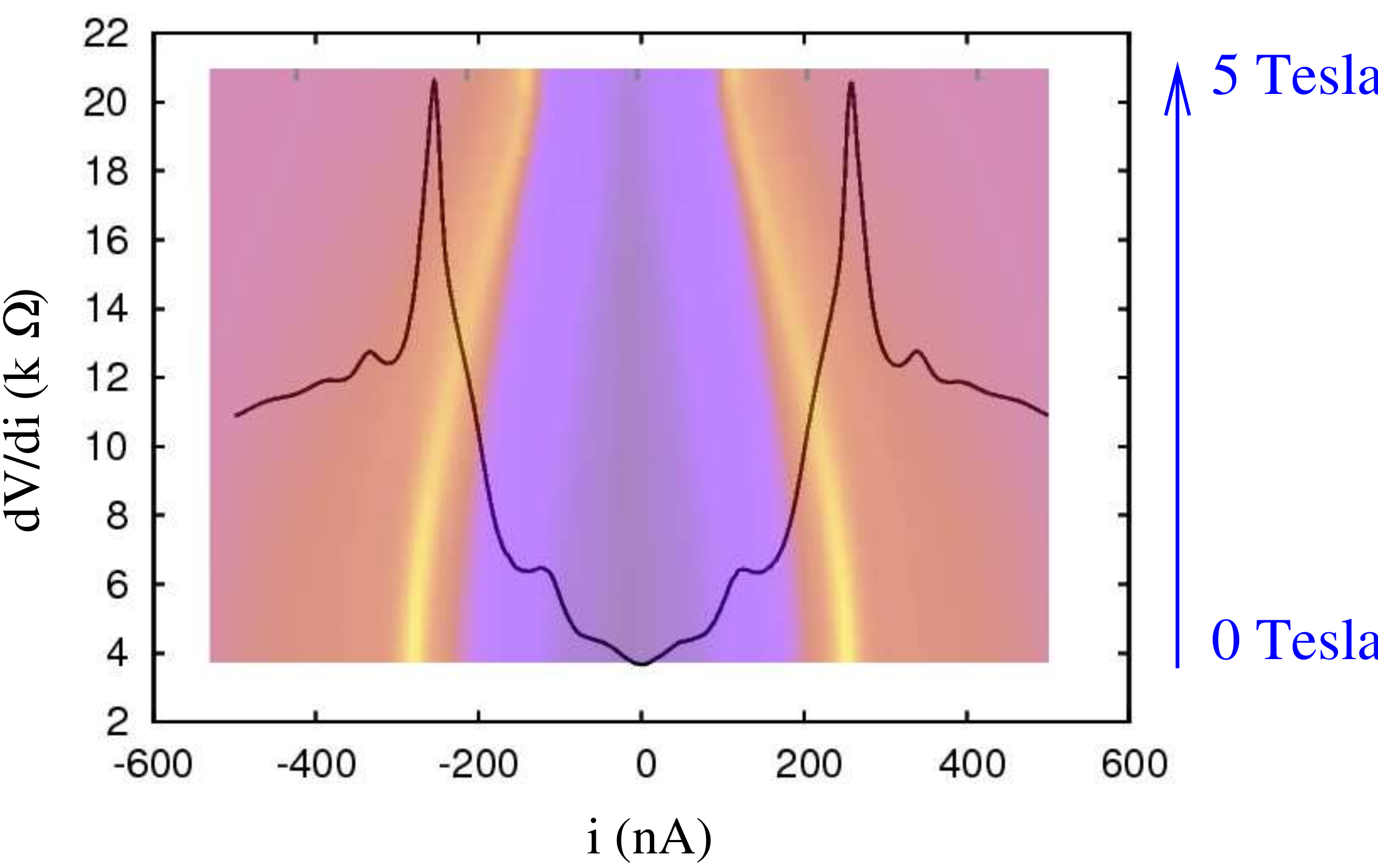}
\caption{The black curve represents the differential resistance $dV/di$ as a function of DC current through the $10\;{\rm k\Omega}$ 
sample at $100\;{\rm mK}$. The color inset in the background shows the evolution of the differential resistance
encoded as a color scale with yellow/violet representing maximal/minimal differential resistance. 
The $x$ axis represents the DC-current as in the main figure, and the $y$ axis indicates the magnetic field 
ranging from $0$ to $5$ Teslas. 
}
\label{fig:FigDnadVdI}
\end{figure}

\begin{figure}[ht]
\centering
\includegraphics[width= 0.8 \columnwidth]{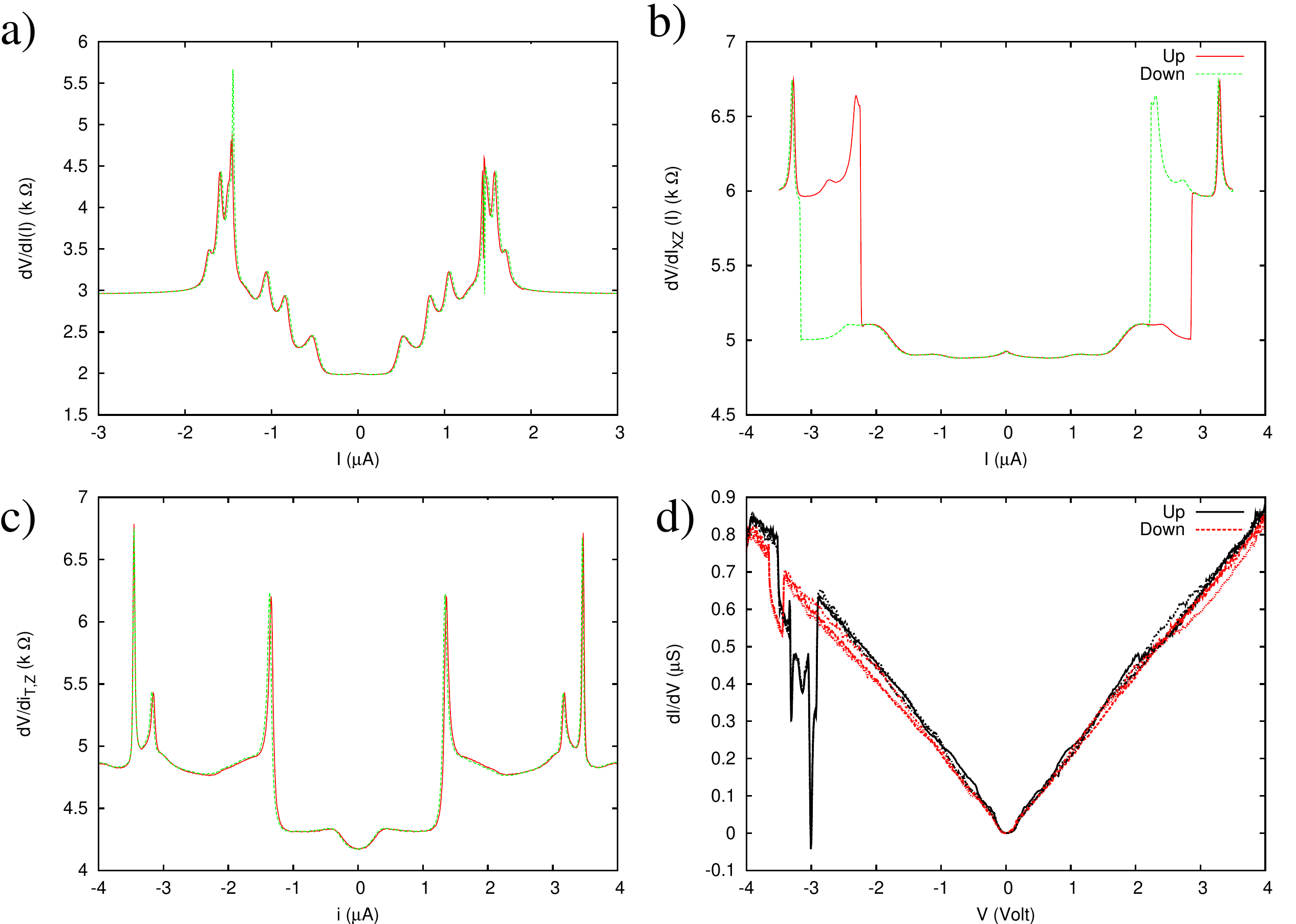}
\caption{The panels a,b,c. show the differential resistance of the $3\;{k \Omega}$, $4.8\;{k \Omega}$ and $6.1\;{\rm k \Omega}$ samples
which have a superconducting behavior. The panel d. displays the differential conductance of the resistive $10\;{\rm M \Omega}$ sample.
Temperature was $100\;{\rm mK}$. 
}
\label{fig:FigDnadVdI4}
\end{figure}

We have also measured the differential resistance of the sample as a function of a DC current.
The results for the $10\;{\rm k\Omega}$ sample are shown on Fig.~\ref{fig:FigDnadVdI} and 
resemble the data obtained from Ref.~\cite{Kasumov} on DNA samples with superconducting Rhenium contacts. 
At zero magnetic field the differential resistance exhibits a large drop in the current interval $-250\;{\rm nA} \le i \le 250\;{\rm nA}$.
Inside the gap region the resistance drops smoothly but does not vanish probably because of the  contact resistance of the normal electrodes. 
In our interpretation the large peak at $i \simeq \pm 250\;{\rm nA}$ corresponds to the critical current of a nanoparticle.
The smaller peaks may appear due to other nanoparticles weakly connected in series or in parallel in our conducting junction.
As suggested by the colorscale diagram, the critical current decreases when a magnetic field is applied. 
At the maximal available magnetic field $H \simeq 5\;{\rm Tesla}$, the critical current is $i \simeq 130\;{\rm nA}$
since as stressed above our magnetic fields are not strong enough to completely suppress superconductivity.
The three other less resistive junctions give similar $dV/di$ dependence (see Fig.~\ref{fig:FigDnadVdI4}).
In these other samples more peaks are apparent and one of the junctions was hysteric. Figure~\ref{fig:FigDnadVdI4} also
presents the low temperature differential conductance of the resistive junction with room temperature 
resistance of 10$\;{\rm M\Omega}$. At low temperature this sample becomes insulating at low bias voltage.
When the bias exceeds $100\;{\rm mV}$ the conductance starts to increase following a cone shape typical 
for graphite but a priori unexpected in our samples. An hysteric singularity appears when the bias reaches a value around $-3\;{\rm V}$.
Although it is hard to determine with certitude the origin of this hysteresis we note that $-3\;{\rm V}$ 
is close to the estimated HOMO-LUMO gap in DNA and that similar singularities were observed at room temperature in the 
conduction of DNA/lipid films confined between nanogaps  \cite{Kasumov3}.  In this respect this is the only
sample whose DC transport characteristics can hardly be mimicked by a metallic short-circuit of very small dimensions. 

\begin{figure}[ht]
\centering
\includegraphics[width=0.65 \columnwidth]{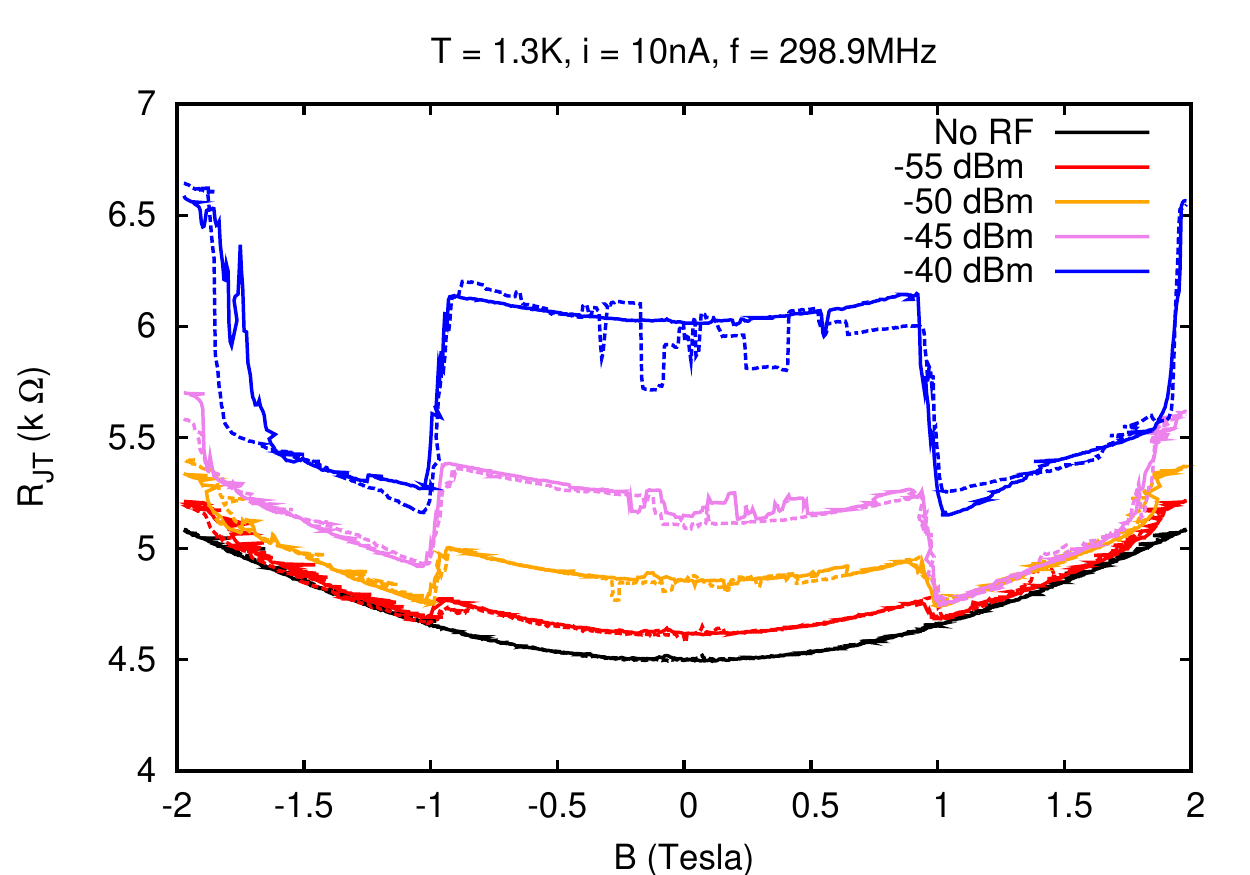}
\caption{Magnetoresistance of the $10\;{k \Omega}$ sample for several microwave powers. Microwave frequency was $f = 298\;{\rm MHz}$. 
}
\label{fig:FigDnaRF}
\end{figure}

In order to search for sample characteristics which might be specific of DNA molecules we have irradiated 
our samples with microwaves. Our idea was that the helix structure of the molecule could induce 
special magnetic field asymmetry in the out of equilibrium transport across the molecule. 
This expectation was not confirmed experimentally since the $R(B)$ dependence under irradiation remained rather symmetrical.
However the DC-magnetoresistance of our samples could become unstable under microwave irradiation (see Fig.~\ref{fig:FigDnaRF}). Interestingly 
instabilities were observed mainly at rather low frequencies $f < 1\;{\rm GHz}$. 
A possible (although science fiction like) interpretation is that the microwave field excites a mechanical transition between
two possible equilibrium positions for a DNA molecule suspended across the peaks created on both sides of the gap by the FIB etching;
in this scenario the superconductivity just enhances the sensibility to these mechanical vibrations.
However one must take into account that the response to microwave may be very complicated in superconducting weak links
where the switching may become chaotic. In particular magnetic field anti-symmetric photovoltaic effect was observed in such systems 
by \cite{Bartolo}. Hence the presence of a magnetic field asymmetry does not allow to discriminate between 
a chiral molecule like DNA and an array of superconducting weak links. 

\begin{figure}[ht]
\centering
\includegraphics[width=0.6 \columnwidth]{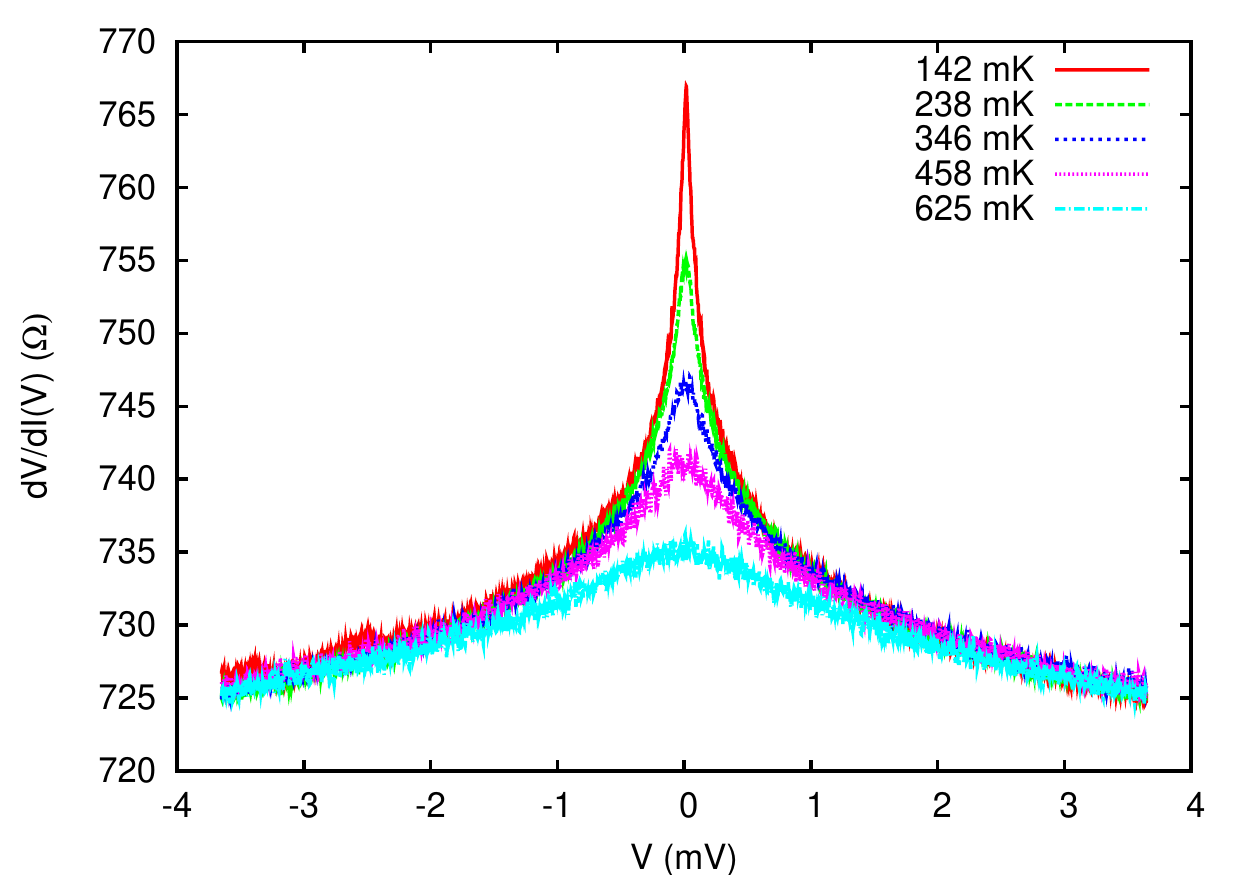}
\caption{Differential resistance at several temperatures as a function of applied voltage for the control short-circuit sample without DNA. 
During the FIB etching the sample resistance was increased from $1.2\;{\rm k\Omega}$ to $1.9\;{\rm k\Omega}$.
}
\label{fig:FigPtNoDNA}
\end{figure}

In the above experiments we have assumed that superconductivity was induced by superconducting nanoparticles because of the high critical magnetic fields. 
Yet it is not the only possibility since FIB irradiation can induce superconductivity in materials with otherwise very small critical temperatures.
For example the tungsten deposited under FIB has critical temperatures around $4\;{\rm K}$ and a high critical magnetic field of $4\;{\rm Tesla}$. 
Hence there is also a possibility that the Pt bumps observed on the edges of the gap (see Fig.~\ref{fig:FibAFM}) could become superconducting 
due to a high concentration of incorporated Gallium atoms. In order to check experimentally if this is indeed the case we have 
deliberately prepared a short-circuit across a clean Pt film by stopping the FIB etching before the gap became completely insulating
(this may be compared with  Fig.~\ref{fig:FibEtchTime} where the metallic film was etched until the gap became insulating).
In this control sample superconductivity was absent, but low temperature measurements revealed a low bias differential resistance 
peak increasing at low temperature (see Fig.~\ref{fig:FigPtNoDNA}) in a behavior reminiscent of dynamical Coulomb blockade \cite{Klinov}.
The effect in Fig.~\ref{fig:FigPtNoDNA} is smaller by an order of magnitude compared to Ref.~\cite{Joyez}.
This is probably because we measured a short-circuit much more conductive than conductance quantum, 
however short-circuits formed by FIB can be more resistive see for e.g. Fig.~\ref{fig:FibEtchTime}.
Hence it is plausible that FIB etching can create quasi-one dimensional filaments inside the gap capable of producing dynamical coulomb blockade behavior.

In general these superconducting nanoparticles are unlikely to induce superconductivity in a normal metallic conductor 
because of the inverse proximity effect. Direct proximity effect is possible in this regime only if the density of states 
in the normal metal is very low. For example it was recently proposed that nanoparticles could induce superconductivity 
in graphene \cite{Kostia}. Since DNA molecules have a very low density of states they are good candidates
for forming the normal region of our SNS  junctions. The image that then comes to mind (see Fig.~\ref{fig:FigPtNoDNA}) is that of a 
DNA molecule connecting several nanoparticles and connected to the normal PtC contacts. 

However it is also possible that the nano-filaments 
created by FIB can be small enough for the direct proximity effect to be possible. 
Indeed, even if the gaps were insulating before deposition of DNA, one can imagine that 
the metallic residues inside the gap could have been displaced by the DNA molecules 
creating a leak. Note that the absence of superconductivity in the 
control sample from Fig.~\ref{fig:FigPtNoDNA} does not rule out this scenario. Indeed we will show in the next 
section that the deposition of nanoparticle is not a reproducible phenomena and a single control short-circuit sample may not 
be representative. It is actually hard to discriminate between these two possibilities. Our samples do not have a backgate
since they are prepared on mica,  heating up to the denaturation temperature of DNA is not possible since pure Gallium melts at low 
temperature of $\simeq 30\;{\rm ^oC}$ and experiments in liquid may simply break the device by removing the pentylamine layer.
As a result we have chosen ultraviolet (UV) irradiation as the most distinctive experiment, indeed it is well established 
that UV can damage the DNA molecules whereas it is not harmful for a metallic film provided that there is no heating 
from the UV lamp. 
We tried to irradiate one of the samples with an UV irradiation with wavelength $\lambda = 233\;{\rm nm}$.
The electrical conduction disappeared immediately (on the scale of seconds) after the UV lamp was switched, 
whereas the resistivity of a platinum film of a few nanometers  thickness did not change after an hour of irradiation.
Even if this experiment is spectacular it is not necessarily conclusive because an electrical discharge could have occurred 
when the lamp was switched on (the lamp and resistance measuring equipment are connected through a common connection to the ground). 
Probably in future experiments UV must be attenuated to observe a more progressive transition.

\begin{figure}[ht]
\centering
\includegraphics[width=0.6 \columnwidth]{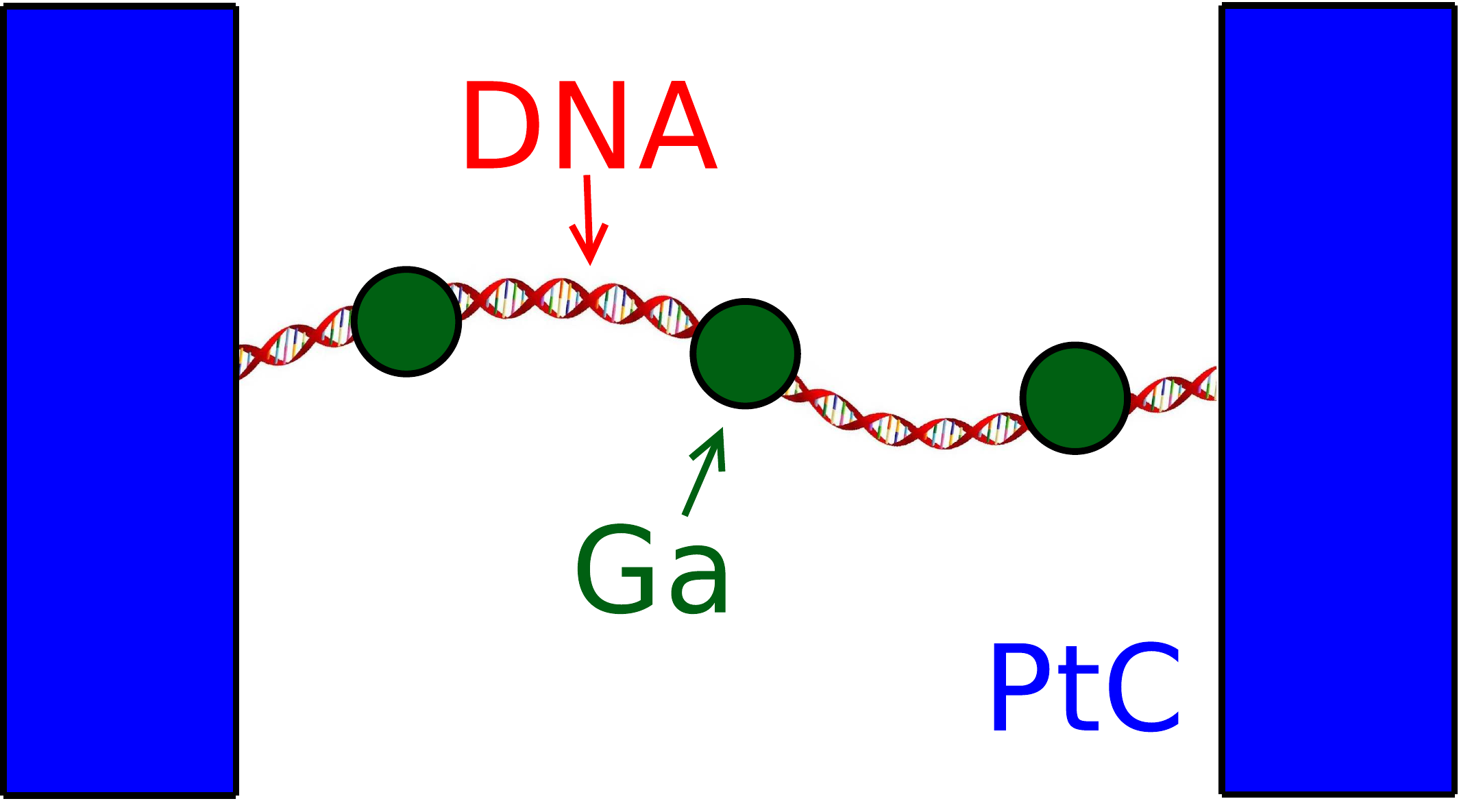}
\caption{Schematic representation of a conducting DNA molecule contacting the normal ${\rm PtC}$ electrodes and superconducting 
Gallium nanoparticles. 
}
\label{fig:DnaDrawNano}
\end{figure}

If our interpretation in term of proximity-induced superconductivity in DNA molecules is valid, 
the number of peaks in the differential resistance measurements (see Fig.~\ref{fig:FigDnadVdI} and Figs.~\ref{fig:FigDnadVdI4}.a,b,c) 
gives an estimation of the number of nanoparticles connected 
to the DNA molecules that transit to the superconducting state.
Two limiting cases may occur depending on the number of connected DNA molecules that exhibit proximity induced superconductivity : 

\begin{itemize}

\item {\bf 1.} It is possible that only a single molecule transits into the superconducting state.
In this case each peak in the differential resistance should correspond to the critical current of a DNA  
molecule connected to superconducting nanoparticles and the number of peaks 
should give the number of  nanoparticles connected to the superconducting DNA molecule. 
In our samples the number of peaks varies from $3$ to $6$ (sample from Fig.~\ref{fig:FigDnadVdI4}.b and Fig.~\ref{fig:FigDnadVdI4}.a respectively).
By dividing the average length of the gap by the number of connected nanoparticles, 
we can estimate the typical length of the individual DNA segments that connect 
neighboring nanoparticles (see Fig.~\ref{fig:DnaDrawNano} for a sketch of the geometry).
For a gap width of approximately $100\;{\rm nm}$, 
we find that this length scale varies from $15\;{\rm nm}$ to $30\;{\rm nm}$. 

\item {\bf 2.} In the opposite limit, we can assume that there are only two connected nanoparticles per superconducting molecule.
Under this assumption the number of peaks gives the number of superconducting molecules.
The transport in DNA molecules in this configuration is typically probed on a length scale corresponding 
to half of the gap width: $50\;{\rm nm}$. 

\end{itemize}

In conclusion the conductivity of DNA molecules is probed on a length scale between $10$ and $50\;{\rm nm}$ 
which is smaller than the width of the insulating gap which is around $100\;{\rm nm}$ wide. We note that transport in DNA on a $10\;{\rm nm}$ scale 
was reported by several independent groups for e.g. \cite{Porath,Barton}.

\clearpage 

\begin{figure}[ht]
\centering
\includegraphics[width=0.7 \columnwidth]{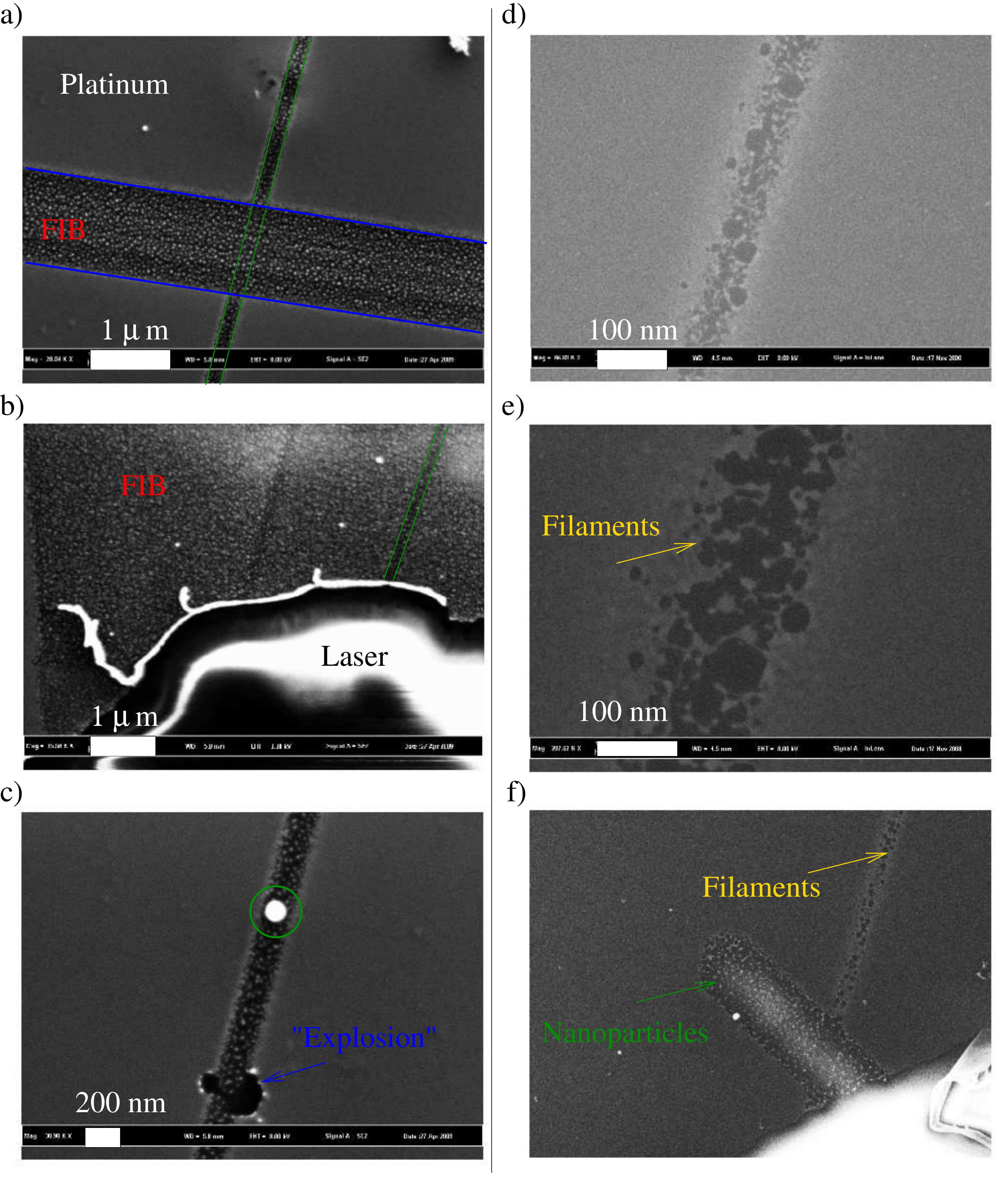}
\caption{(left) SEM images of the gaps from our experiment. Images a,b and c were taken on the control sample from Fig.~\ref{fig:FigPtNoDNA}. 
Even if we know that a short-circuit was present we could not find it under SEM, probably because the slit is long and the short-circuit is very narrow. 
(right) The SEM images d,e and f where taken from a sample used for deposition of DNA molecules. }
\label{fig:DnaSem}
\end{figure}

\clearpage 

\section{Characterization of the gaps from our transport experiments}
\label{chap:FibAFMSEM}

The topography characterization of the gaps etched by FIB was already exposed in Section~\ref{chap:FIB}
and we will start this characterization section by describing the scanning electron microscope (SEM) images of our gaps. 
SEM provides a strong contrast between insulators and metals, whereas AFM in tapping mode is only sensible to topography. 
Hence SEM is a good technique to detect metallic residues inside the gap where the topography is rough after FIB etching.
Note that no careful SEM analysis of the gaps was done in the previous experiments \cite{Kasumov,Klinov}, 
however it is very likely that the gaps used in these experiments had a similar structure specially in \cite{Klinov}
where they were also fabricated by FIB. 

Several SEM images of the FIB gaps are displayed on Fig.~\ref{fig:DnaSem}, 
intriguingly in all the images metallic contamination is present in the gap after etching. 
Two possible limit situations seem to emerge from our observations. In the case 
where the FIB dose is weak the metallic residues take the form of filaments (Fig.~\ref{fig:DnaSem} right panel) 
whereas after a stronger dose the residues seem to form an ordered network of nanoparticles (Fig.~\ref{fig:DnaSem} left panel).
This transition may be inferred from  Fig.~\ref{fig:DnaSem}.f, where two neighbor regions were etched with 
different FIB doses under the same beam conditions. In the narrow trench filaments can be clearly distinguished 
(see also Figs.~\ref{fig:DnaSem}.d,e ) while in the wider trench where the dose was stronger residues form individual nanoparticles. 
Naturally one can expect the residues to disappear after a sufficient FIB dose. This dose however seems difficult 
to attain in practice. This difficulty is illustrated on Figs.~\ref{fig:DnaSem}.a,b. In the first figure two 
intersecting regions were etched with FIB (they are highlighted by blue and green lines). 
Both regions were exposed to a dose sufficient to etch most of the metallic film leaving isolated nanoparticles inside the gap. 
Surprisingly the density of nanoparticles does not decrease in the intersection between these two regions (parallelogram with 
edges formed by blue and green lines) even if the intersection area received a dose about two times larger than the other regions. 
A similar situation is observed on Fig.~\ref{fig:DnaSem}.b, where a large area was etched by FIB irradiation around a 
slit cut by FIB (delimited by the parallel green lines). The region inside the slit still contains nanoparticles
even if it received twice an irradiation dose capable of etching most of the platinum film. 
As a result the dose needed to completely etch all metallic residues,
is certainly much larger than the minimal dose required to create an insulating slit.

\begin{figure}[ht]
\centering
\includegraphics[width=0.7 \columnwidth]{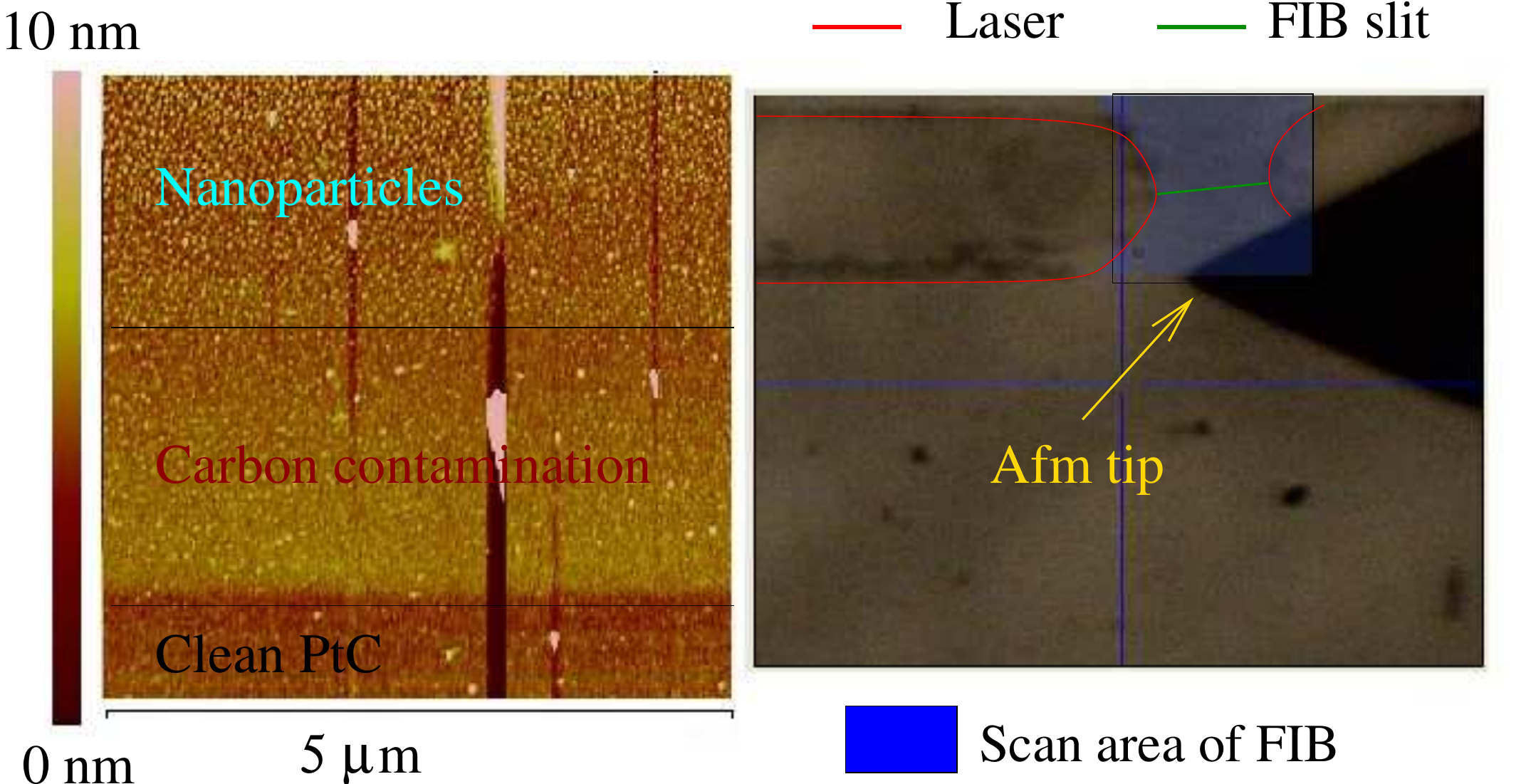}
\caption{(left) AFM image showing the boundary of the area contaminated by nanoparticle, it almost coincides with 
the border of the contamination layer deposited by FIB. This image was taken on the sample measured at low temperature in Section~\ref{chap:DnaTransport} and is representative of the other slits on that substrate. The typical height of the nanoparticle is between $5$ and $7\;{\rm nm}$ which 
is compatible with the estimates from Section~\ref{chap:DnaTransport}.
(right) The position of the AFM tip is shown on an optical microscope photograph and coincides with the border of the FIB scan
windows with size $\simeq 100\;{\rm \mu m}\times 100\;{\rm \mu m}$). }
\label{fig:NanoPart}
\end{figure}

The presence of nanoparticles has two possible origins, an instability of the gallium tip that can start 
to emit small nanoparticles instead of individual gallium ions and the recombination of the gallium atoms 
on the mica surface (the latter mechanism is then specific to mica since it is not observed on ${\rm Si/SiO_2}$). Our observations suggests that 
both mechanisms are possible. On one sample we have detected nanoparticles everywhere inside 
a large square area corresponding to the scan size of the FIB (see Fig.~\ref{fig:NanoPart}). 
This proves that in some uncontrolled regime FIB is capable of disseminating small nanoparticle instead of individual gallium ions.
The above behavior was observed on only a single sample, however this was the sample that we studied at low temperature  
in Section \ref{chap:DnaTransport}, hence this rather special case is very important for the interpretation of our transport measurements. 
In other samples nanoparticles were observed only inside the gap indicating a different origin. 
It may be possible that Gallium atoms recombine on the mica surface to form nanoparticles.
Note that a similar behavior has been observed on GaAs substrates \cite{OrsayPhysics} and the presence of Aluminum atoms in mica 
substrate may play a role. 

The atomic force microscopy measurements on the edge of the FIB scan window depicted on Fig.~\ref{fig:NanoPart}, 
revealed another source of contamination. A step of $1-2\;{\rm nm}$ height systematically surrounded 
the FIB scan window, this contamination layer is probably composed of carbon deposited during FIB imaging of
the gaps (similarly to a contamination layer deposited under SEM). This is actually very 
discouraging because the FIB technology was chosen because it supposedly limited contamination.
In reality the electrodes are covered by a poorly characterized carbon contamination layer.
As a side consequence this makes the debate on platinum versus platinum carbon electrodes not pertinent near the slits. 

We will now summarize the different contamination sources that were revealed through SEM and tapping mode AFM microscopy. 
On mica substrate it seems that metal residues are always present inside the gap (at least for practically reasonable 
FIB doses). They can take the form of narrow filaments for low FIB doses or a net of nanoparticles at high doses. 
An exceptional regime can also occur where the gallium tip of the FIB becomes unstable and starts to stew nanoparticles 
everywhere in its scan area. From our experiments this case seems rather improbable however it occurred at least once,
on the sample that we measured at low temperatures. Finally a carbon contamination layer is deposited near 
the gap during the imaging inside the FIB microscope. Keeping in mind the above information, we will now 
focus on the detection of DNA molecules across the gaps where conduction was observed after deposition of DNA molecules. 

In previous experiments \cite{Kasumov,Klinov} DNA molecules crossing FIB slits could be detected with tapping mode AFM.
However AFM and low temperature transport measurements were done independently, and probably on different samples 
while we know that  fluctuations are very strong from one sample to another. 
Thus the correlation between the presence of DNA molecule and the appearance of electronic transport was supported only by a statistical argument
(absence of conduction on control samples when a buffer solution without DNA was deposited).
Here we tried to detect DNA molecules on the three samples where conduction was induced after the deposition of  DNA 
molecules with the pentylamine technique was attempted.

\begin{figure}[ht]
\centering
\includegraphics[width=0.8 \columnwidth]{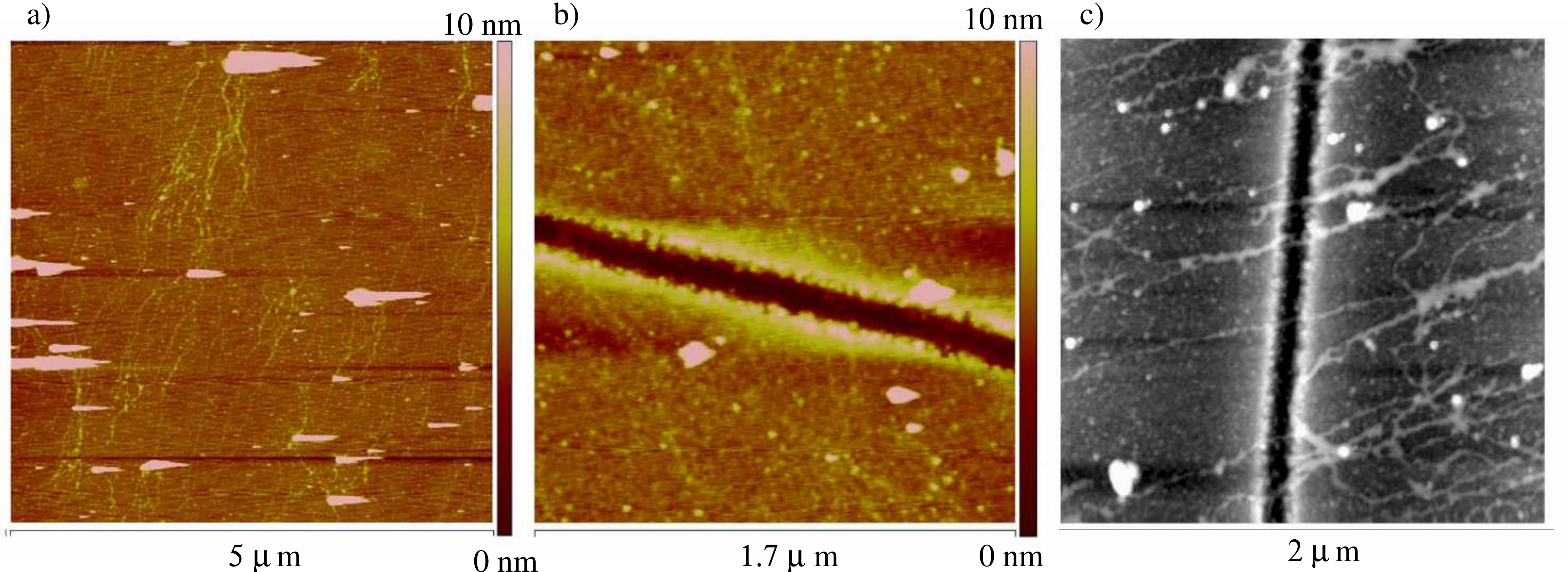}
\caption{(Sample DnaBox3B) AFM microscopy images of a sample where conduction was observed after deposition of DNA. 
DNA molecules can be observed far from the gap on the clean PtC surface (panel a) and near the gaps.
Some molecules crossing the slits were also observed on this same sample (panel b).
The gap shown on panel b was conducting after deposition of DNA and took the following resistances 
$1.8\;{\rm k \Omega} \rightarrow 160\;{\rm k \Omega} \rightarrow > 1000\;{\rm k \Omega} \rightarrow 5\;{\rm k \Omega} \rightarrow \infty$
while we tried to make contacts with silver paint on the sample. 
Panel c. shows an image of DNA molecules crossing a slit obtained by D.Klinov, most likely on the same sample.
\label{fig:Box3B}
}
\end{figure}

We will start with two samples where conduction was destroyed during the attempts to contact the samples electrically (see Section\ref{chap:DnaTransport}).
In one of the samples DNA  molecules could be observed both in the region far from the gaps where PtC was clean 
and inside the area covered by the carbon contamination film near the gap as illustrated on Fig.~\ref{fig:Box3B}.
On this sample we could confirm the presence of DNA on some of the conducting slits. It was however impossible to 
make a complete statistic  because after the electrical conduction was destroyed we tried to recover the conduction by keeping 
the sample in a humid atmosphere for a couple of days. This procedure did not restore the conductivity but it led to the formation 
of unidentified ``contamination pancakes'' on the sample surface  which impeded further AFM characterization.

\begin{figure}[ht]
\centering
\includegraphics[width=0.8 \columnwidth]{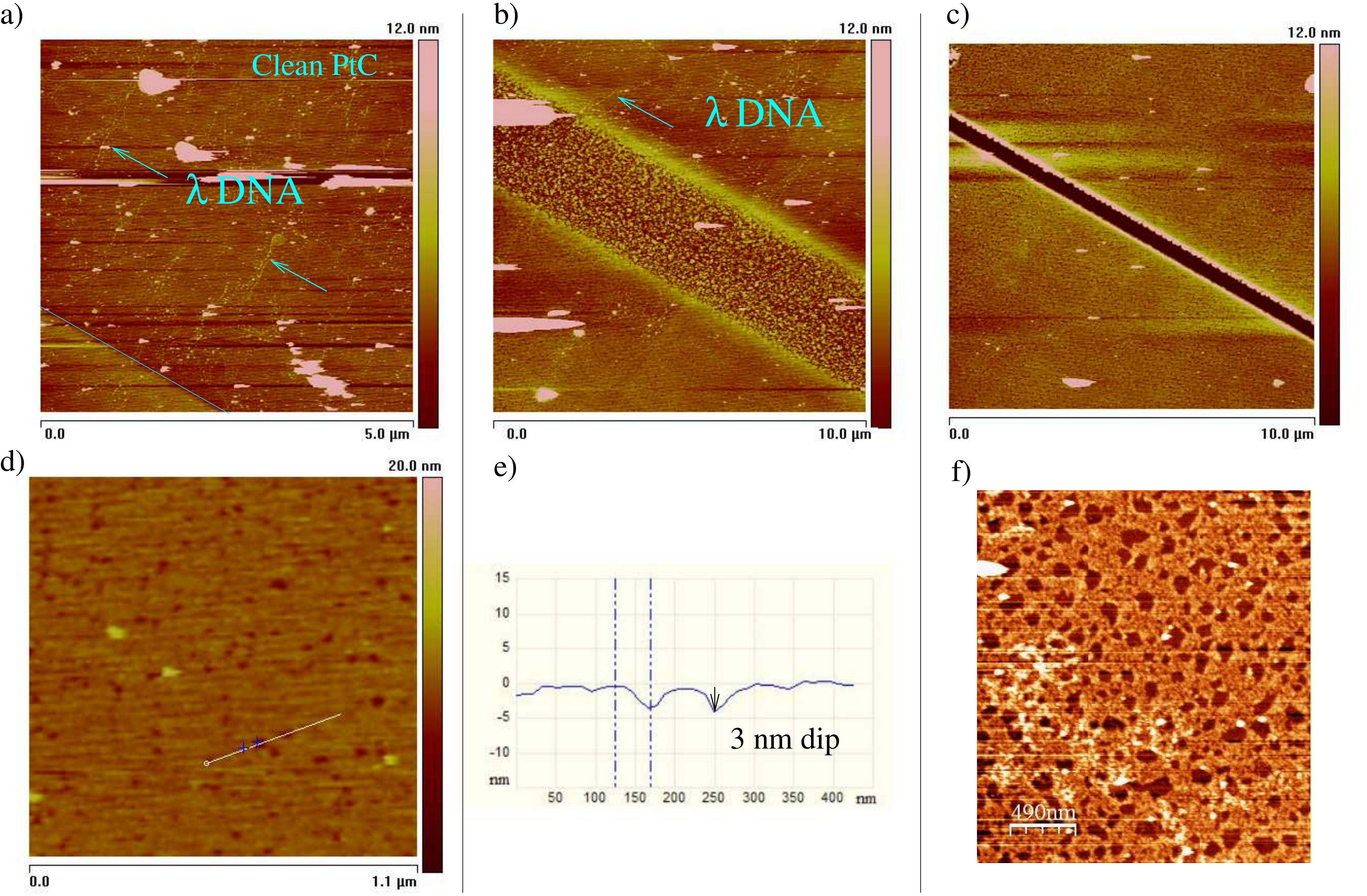}
\caption{(Sample DnaBox3A) AFM microscopy images of one of a sample where conduction was observed after deposition of DNA molecules. 
As on Fig.~\ref{fig:Box3B} DNA molecules can be observed far from the gap on the clean PtC surface (panel a). 
some DNA molecules cross the border of the carbon contamination. 
On panel b) the $\lambda$ molecules cross a large trench (not connected electrically) cut by FIB inside the carbon contamination layer. 
The $\lambda$ concentration seems to drop on the bottom side of the trench, one can notice small holes in the substrate in this region. 
Panel c) is an AFM scan around the narrow FIB slit that is connected electrically. No DNA molecules could be seen in this region,
although we were able to resolve many DNA molecules on the same scale on panels a) and b). Panel d) shows in more detail 
the small dips in the surface. The depth of these holes is measured on panel e) and is approximately  $\simeq 3\;{\rm nm}$. 
Panel f) displays another sample where holes appeared after deposition of DNA on a PtC substrate using 
pentylamine functionalization (hole depth was $\simeq 1\;{\rm nm}$). The formation of holes was not observed on other samples.
\label{fig:Box3A}
}
\end{figure}

In the next samples (DnaBox3A) were taken after silver paint and ultrasound bonding that
destroyed conductivity across the slits. The experiments were done in this order because 
priority was given to transport measurements over AFM characterization and we were afraid that 
the AFM tip could damage conducting DNA molecules when scanning. 
Hence only the previous sample (Fig.~\ref{fig:Box3B}, DnaBox3B), was well characterized before 
transport measurements. However the fluctuations between the samples were strong and Fig.~\ref{fig:Box3A} 
revealed a different behavior from sample DnaBox3B. 
Many combed $\lambda$ molecules could be observed far from the thin FIB slits outside the carbon contamination layer.
Near the slits however the surface was very different, it included small holes around 3 nanometers deep 
and no visible DNA molecules.

\begin{figure}[ht]
\centering
\includegraphics[width=0.8 \columnwidth]{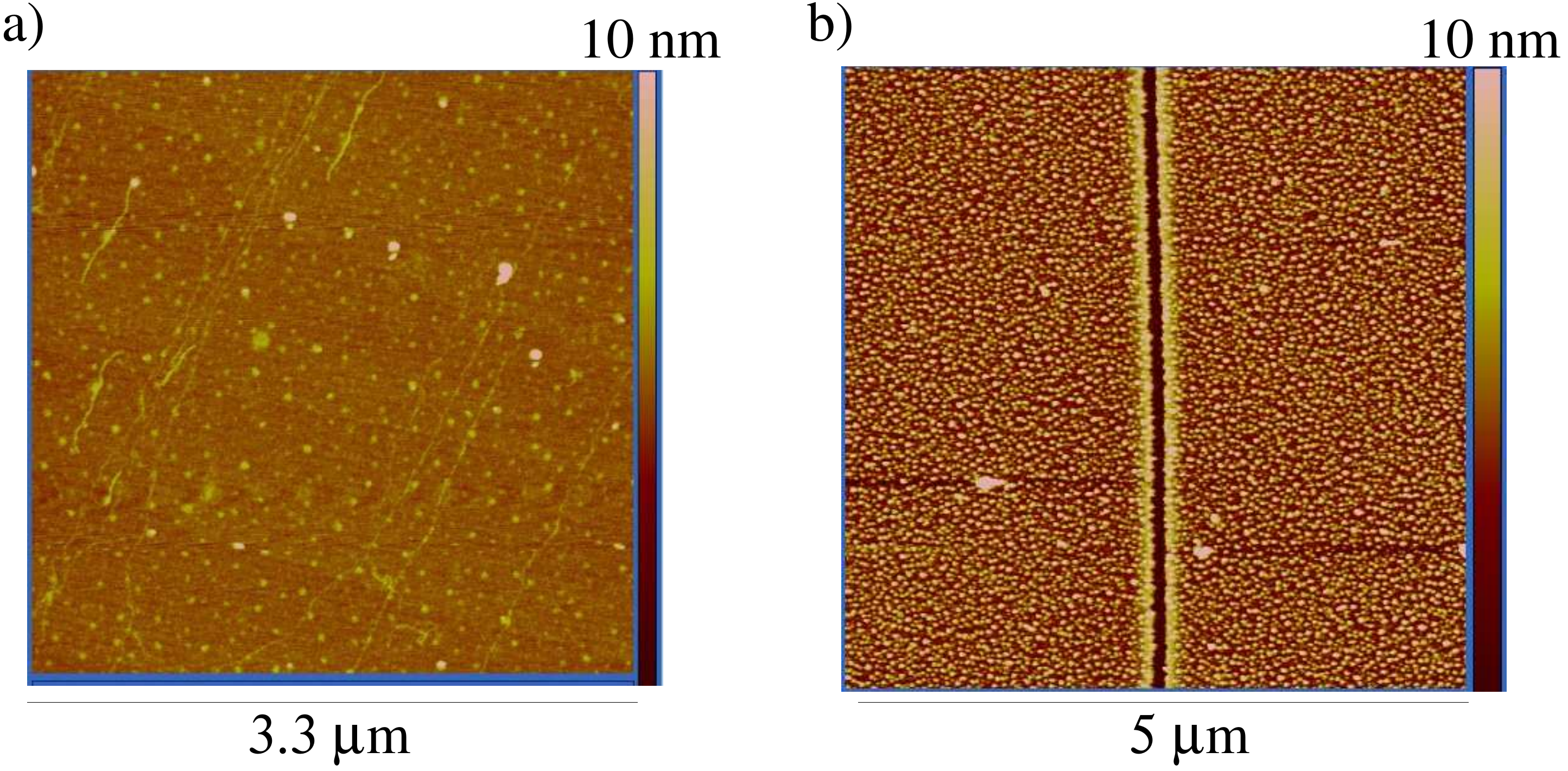}
\caption{(Sample DnaBox2B) AFM microscopy images from the sample where low temperature transport was investigated in Section~\ref{chap:DnaTransport}. 
Panel a) shows DNA molecules observed far from the slit, panel b) shows a $5\;{\rm \mu m m}\times5\;{\rm \mu m}$ image of
a slit that became conducting after deposition of $\lambda$-DNA molecules. The sample surface around the slit is 
covered by nanoparticles which makes imaging molecules difficult. Fig.~\ref{fig:Box2Bdima} shows AFM images  
obtained with an ultrasharp AFM tip that allow to resolve DNA molecules across the slits.
\label{fig:Box2B}
}
\end{figure}

\begin{figure}[ht]
\centering
\includegraphics[width=0.8 \columnwidth]{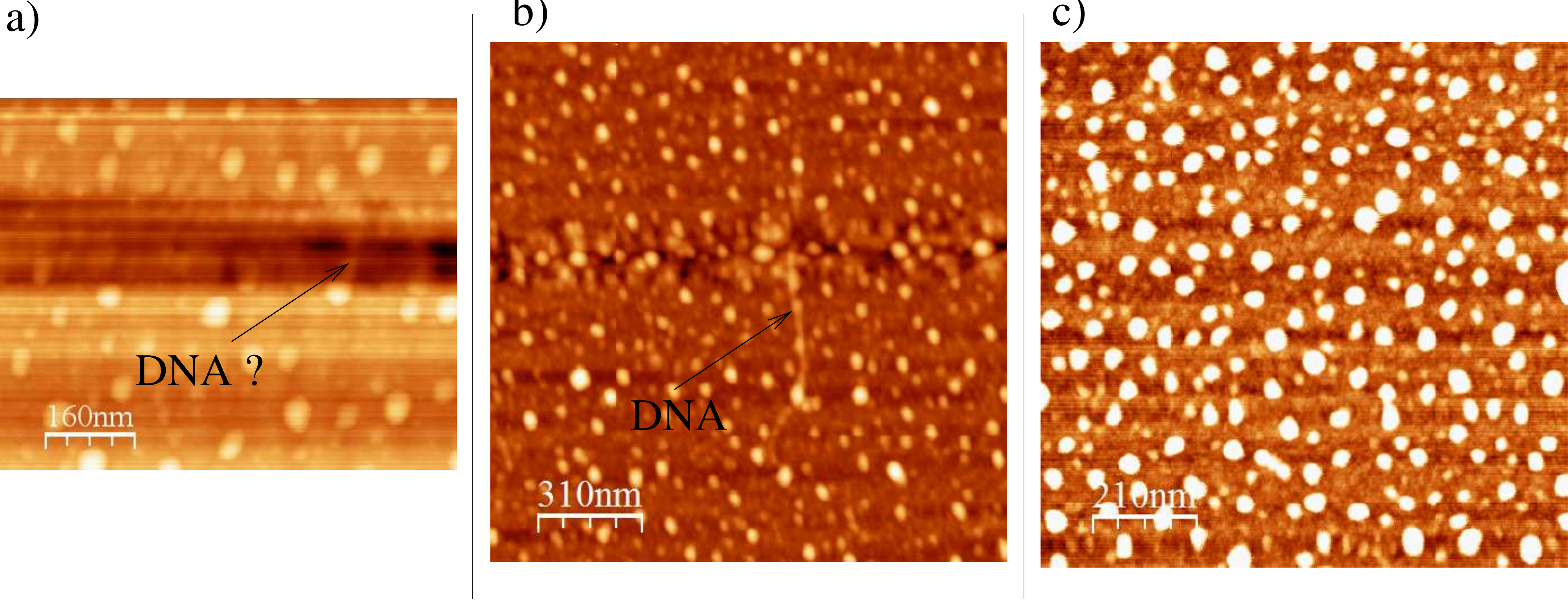}
\caption{(Sample DnaBox2B) AFM microscopy images from the sample where low temperature transport was investigated in Section~\ref{chap:DnaTransport}. 
These images were taken by Dmitry Klinov using an ultrasharp AFM tip, some images show the presence of DNA molecules close to the insulating gaps (panel a) and b)).
However on typical images DNA molecules seem absent  (see panel c of this figure, or Fig.~\ref{fig:Box2B} which was obtained in Orsay).
\label{fig:Box2Bdima}
}
\end{figure}

The absence of DNA molecules near the gap inside the carbon contamination layer, 
was observed for at least two other samples, including the sample DnaBox2B where electron transport
was investigated in Section~\ref{chap:DnaTransport}. The situation on this sample is depicted on Fig.~\ref{fig:Box2B},
far from the gap the surface is clean. The region near the slit is covered with nanoparticles (see also Fig.~\ref{fig:NanoPart}) 
but the apparent density of DNA molecules seems very low. 
On the one hand the fact that we do not observe DNA molecules does not prove that they are completely absent in the gap. 
Imaging conditions could be spoiled because of the presence of nanoparticles/holes in the contamination layer.
We note that further AFM characterization, performed by Dmitry Klinov using a sharper tip revealed some molecules 
crossing the FIB slits (see Fig.~\ref{fig:Box2Bdima}), which supports this hypothesis.
On the other hand, it is also possible that the presence of a FIB contamination layer reduces significantly the efficiency
of the pentylamine plasma. This claim may seem strange since the experiments described in Section~\ref{chap:dnacomb} demonstrate that pentylamine is 
a reliable technique to attach DNA on metallic electrodes with a sufficient quantity of carbon atoms on the surface. 
The above description apriori apply to the FIB contamination layer. A hint on why this argument can fail,
is provided by the insitu transport measurements inside the FIB chamber. As mentioned in Section~\ref{chap:FIB} 
the resistance generally decreased slightly at the beginning of the FIB etching. Probably during this time the deposition of contamination 
prevailed over etching, the drop of resistivity then indicates that the carbon contamination layer is rather conducting 
and probably locally graphitic. This claim is supported by the fact that platinum can be used as a chemical vapor deposition (CVD)
catalyst for the growth of single wall carbon nanotubes \cite{PtSWNT}. A last argument in this direction comes from the differential 
conductance of the resistive sample on Fig.~\ref{fig:FigDnadVdI4} which has a cone structure very similar to the density 
of states in graphene; it is possible that accidentally the current passed through a flake of few layer graphene in this sample. 
Since graphite is very stable the available number 
of carbon atoms that can serve to anchor the pentylamine plasma can be greatly reduced compared to the situation where
for example, the surface is covered with amorphous carbon. This can make the pentylamine film less stable on the surface 
so that sometimes it is removed when the sample is dried in a scenario similar to that described for mica in Section~\ref{chap:dnacomb}.
The holes observed on Fig.~\ref{fig:Box3A} give a cue in this direction since they show that a layer at least $3\;{\rm nm}$ thick 
was partially removed from the surface. 

To summarize even if DNA molecules were detected across some conducting slits, it was not possible 
to establish a clear correlation between conductivity and the presence of DNA molecules on the basis 
of our AFM measurements. The major obstacle comes from the contamination 
layer deposited by FIB around the slits. Due to this layer DNA deposition near the gap does not take place in the 
same conditions as everywhere else on the clean PtC surface and the density of DNA molecules 
seems reduced near the gap on several samples. We have argued that the pentylamine layer 
may not be stable enough because the carbon layer deposited by FIB is partially graphitic and provides 
few fixation points. This hypothesis also gives an interpretation for the transport data on the 
resistive sample from Section~\ref{chap:DnaTransport} (Fig.~\ref{fig:FigDnadVdI4}.d) where 
a graphene/DNA junction may incidentally have been formed.

\begin{figure}[ht]
\centering
\includegraphics[width=0.7 \columnwidth]{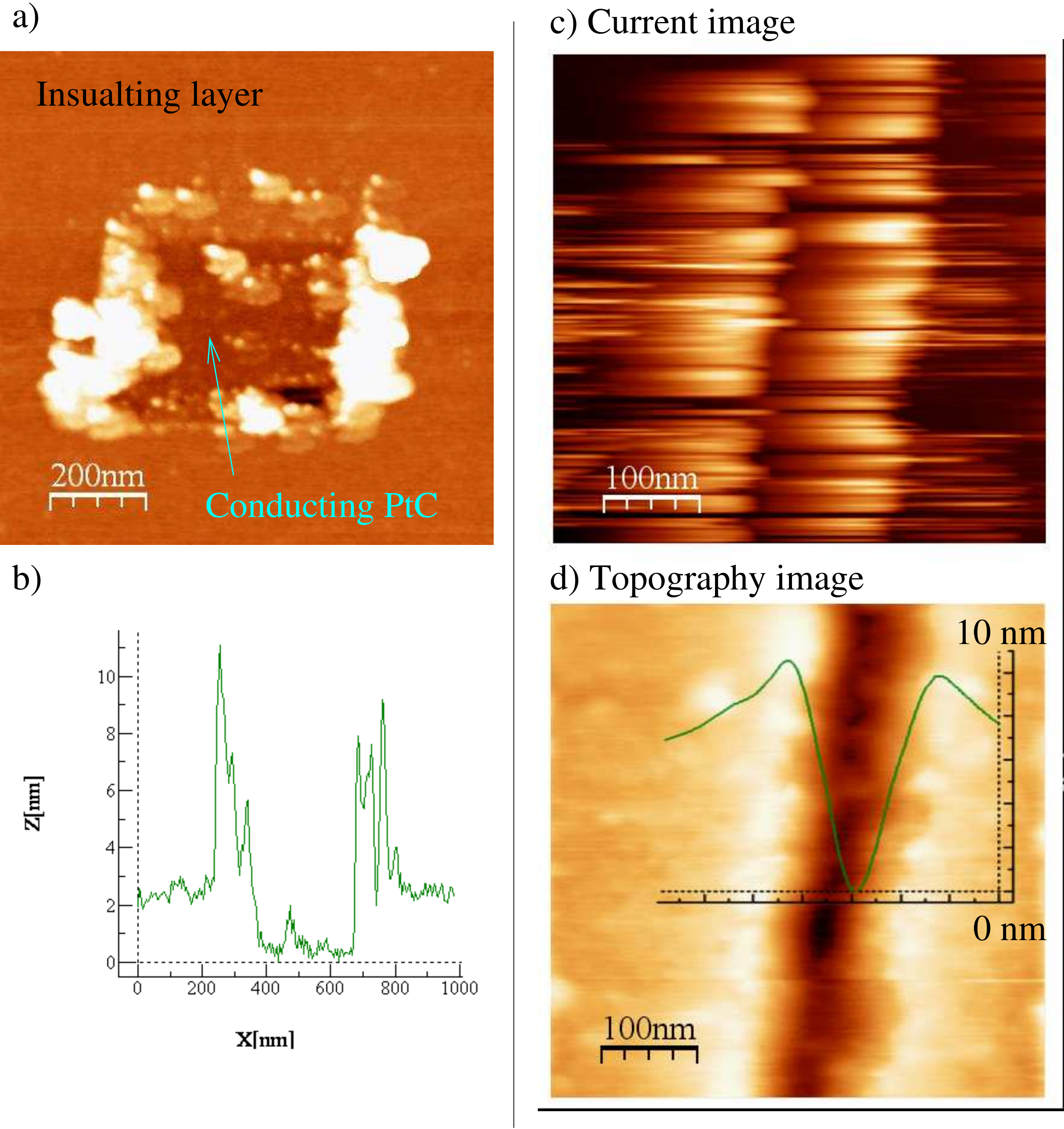}
\caption{
a) The sample is PtC with deposited pentylamine and DNA. 
A small square region was scratched with the AFM tip until conduction was observed. This image is a tapping mode picture 
of the hole formed after this procedure (see cross section on panel b).
c) Conducting AFM image of a conducting slit from Section~\ref{chap:DnaTransport} where low temperature 
transport was studied (see also Fig.~\ref{fig:Box2B}). d) Contact mode topography image of the same slit. 
\label{fig:Lille}
}
\end{figure}

We now turn our attention to the last characterization issue of this section. 
The pentylamine plasma creates a positively charged organic layer that allows to bind DNA molecules onto the electrodes. 
This layer is believed to permit a favorable interaction regime between molecules and the substrate 
where DNA can be conducting \cite{Klinov}. However if this layer is continuous it forms 
an insulating separation between DNA on top of the layer and electrodes at the bottom.
In a few samples we have noticed holes after deposition of DNA on the pentylamine treated surface.
Nevertheless in the majority of cases the surface remained smooth even after deposition of pentylamine and DNA 
molecules with a roughness similar to that of the clean Platinum/Carbon surface. 
We suspect that holes appear when pentylamine is not stable enough on the surface 
and partially desorbs in the buffer during the deposition of DNA. 
From this point of view the depth of the holes gives an estimation of the pentylamine film thickness.
In the two cases displayed on Fig.~\ref{fig:Box3A}.d and Fig.~\ref{fig:Box3A}.f this depth 
is respectively 3 and 1\;{\rm nm}. These values support the picture of a continuous insulating 
layer and cast doubt on the possibility of an electrical contact between DNA and electrodes in presence 
of pentylamine. Note that these values in the range of a few nanometers, 
are even smaller than those measured in Ref.~\cite{Lvov}
where a different plasma functionalization was used to deposit a positively charged organic layer. 
Plasma discharges are even used industrially to grow silicon nitrate when both ammonia and silane 
are present in the discharge chamber with a typical growth rate of $20-50\;{\rm nm/s}$ \cite{Rossnagel,Nabok}.
In view of the above data, our estimates for the thickness of the pentlyamine film are very plausible.

In order to determine experimentally whether electrical contact between DNA and electrodes is possible
in our samples we have carried on conducting AFM experiments at the 
Institute for Electronics, Microelectronics and Nanotechnology (Lille, France) in the group of D. Vuillaume. 
When we performed conducting AFM scans over the PtC electrodes no current was detected between 
the tip with a typical bias of $100\;{\rm mV}$ and the grounded sample electrodes. 
The absence of current is probably due to the insulating pentylamine layer between the tip and the electrode surface. 
To check whether the absence of conduction was due to the insulating pentylamine layer, we scratched the surface in contact mode until conduction 
appeared probably removing the pentylamine  on top of the electrode surface. 
Once a finite current appeared we switched back to tapping AFM mode and scanned the surface over a wider area.
An AFM image obtained during the described experiment is presented on Fig.~\ref{fig:Lille}.b.
A hole roughly $1.5\;{\rm nm}$ deep was drilled in the region where the substrate was scratched until conductivity appeared. 
Probably this value provides a measurement of the pentylamine film thickness which in this case also 
is in the range of a few nanometers.

Conducting AFM data in the bulk of the electrodes indicate that it is very difficult to 
establish an electrical contact between DNA molecules and the electrodes beneath 
probably because of the presence of the pentylamine layer. On the edge of the FIB slits 
however the situation is different. A comparison between conduction (Fig.~\ref{fig:Lille}.c) and topography 
(Fig.~\ref{fig:Lille}.d) images in the region of a slit shows that the edge of the gap 
are not covered by the insulating film and contact is possible in this region. 
The above images were recorded on a slit from the sample where low temperature conduction
properties were investigated in Section~\ref{chap:DnaTransport}.
They also give a clue on why inverse proximity effect does not occur in the conduction model 
from Fig.~\ref{fig:DnaDrawNano} where DNA becomes superconducting due to nanoparticles of diameter $10\;{\rm nm}$
although it is deposited on normal electrodes.
Indeed even if the molecule rests almost entirely over the normal electrodes,
electrical contact can be established only in a region of a few nanometers near the gap.

\clearpage

\section{Conclusions}
\label{chap:DnaConclusions}

We first summarize the experimental results obtained in the previous sections.
\begin{itemize}

\item {\bf Section~\ref{chap:Dnamica}} describes the $\lambda$ DNA solution that was used in 
all our DNA deposition experiments and showed that it yields satisfactory deposition onto mica. 

\item {\bf Section~\ref{chap:metal}} shows that binding through DNA extremities allows to 
deposit molecules on a metallic platinum substrate. No conduction was observed in this
case, in agreement with most of other works. 

\item {\bf Section~\ref{chap:pentylamine}} describes the pentylamine plasma functionalization for 
the adsorption of DNA molecules. We argued that this technique is reliable only on carbon 
coated surfaces. When it is applied on mica the pentylamine layer is probably removed at 
the last step when the sample is dried explaining the absence of 
DNA molecules on the surface. 

\item {\bf Section~\ref{chap:dnacomb}} we demonstrated that reliable combing of DNA molecules across 
insulating slits can be achieved combining pentylamine and rinsing in a steady  flow of water. 
However the samples we prepared with this technique exhibited insulating behavior
for both samples prepared by electron beam lithography and ion beam etching.
A possible reason for the absence of conductivity may be that a thin layer of amorphous 
carbon was evaporated at the last step just before deposition of pentylamine and DNA molecules. 
While it ensured a good adhesion of pentylamine it could render the edges of the 
gap insulating by stabilizing the pentylamine even there (see Section\ref{chap:FibAFMSEM}).
Conducting atomic force microscopy (AFM) measurements could have shown if this explanation is true.

\item {\bf Section~\ref{chap:FIB}} explains how the gaps for deposition of DNA were fabricated
using in situ resistance measurements inside the focused ion beam microscope (FIB).
AFM topography of the gaps is also discussed. Starting from this section deposition of 
DNA molecules was done by D. Klinov. 

\item {\bf Section~\ref{chap:DnaTransport}} describes electrical transport measurements in the low 
temperature regime on samples where conduction was established after the deposition of DNA molecules. 
On four samples superconductivity was observed whereas a last resistive sample had a differential 
conductance similar to graphene. Due to high critical magnetic fields around $10\;{\rm Tesla}$ we interpreted the 
observed superconductivity as proximity effect from superconducting nanoparticles inside the FIB slit. 
On a control sample where a short circuit was formed by stopping FIB etching before 
the sample became insulating no superconductivity was observed. However a zero bias anomaly 
was observed, possibly due to the formation of conducting nanofilaments in the gap during the FIB etching.
Scanning electron microscope images of the filaments that may be formed are shown on Fig.~\ref{fig:DnaSem}.

\item {\bf Section~\ref{chap:FibAFMSEM}} showed scanning electron microscope (SEM) and AFM images of the 
gaps. Metallic nanoparticles were identified on the sample from Section~\ref{chap:DnaTransport},
they appeared because FIB worked in a special regime on this sample where it disseminated Ga nanoparticles everywhere. 
On other samples, both metallic filaments and nanoparticles were observed inside the gap. 
AFM characterization revealed that a carbon contamination layer was deposited by FIB around the gap. 
We conjectured that this contamination layer was partially graphitic, this probably explains the poor reproducibility 
of deposition near the slits while good reproducibility was achieved in Section~\ref{chap:dnacomb}. 
This conjecture also gives insight on the origin of the ``graphitic'' sample that we measured in Section~\ref{chap:DnaTransport}. 
Due to the limits of AFM resolution on the rough surface of the electrodes, 
we could not reach a definite conclusion on the presence of DNA molecules around the gaps. 
Figure ~\ref{fig:Box2Bdima} shows a few molecules in this region, however DNA was absent in most of our AFM images.
Finally both tapping and conducting mode AFM indicate that pentylamine forms an insulating film of a few nanometers 
on top of the electrode surface. The electrical contact between DNA and electrodes seems possible only at the edges 
of the slit were conduction was observed in conducting mode AFM. 

\end{itemize}

In conclusion, several arguments can be retained to demonstrate that 
long range transport across DNA molecules was observed in our experiments. The first argument is statistical,
for transport was not observed after deposition of a buffer solution without DNA. 
However one must be cautious with statistical arguments in these systems where 
sample to sample fluctuations are large. 
We note that the conduction in our samples was systematically destroyed when attempts to bond the samples were made
even when silver paint was used to avoid uncontrolled voltage spikes across the slits.
This observation would be difficult to explain if conduction occured through metalic fillaments only, 
and suggests a denaturation of the DNA molecules by the solvent from the silver paint.
Conduction was also destroyed by UV irradiation with wavelength $233\;{\rm nm}$,
however this experiment was performed on only a single sample and more statistics and better control of irradiation doses are needed. 
Concerning the low temperature transport data, the observed proximity effect suggests that transport 
takes place across a nanowire with a very small density of states. 
It is tempting to conclude from this argument that transport indeed takes place along DNA molecules. 
However here also caution is required since we have shown that FIB can create 
narrow conducting filaments inside the slits whose properties are not well characterized 
(for e.g. they seem to exhibit dynamical coulomb blockade).
Ultimately one must keep in mind that even if the gaps are about $100\;{\rm nm}$ wide on average, 
transport may actually be probed on a much shorter length-scales around $10\;{\rm nm}$ 
due to the presence of metallic residues. 
To summarize our experiments provide indications that long range transport in DNA molecules 
can be achieved through interaction with a disconnected array of metallic nanoparticles, 
however more systematic investigations are needed to determine the largest lengthscales 
on which transport can be achieved in this way. 

\section*{Acknowledgments}
 
We thank F. Livolant, A. Leforestier, D. Vuillaume and D. Deresmes for fruitful discussions. 
We acknowledge support from ANR QuandADN and DGA. One of us, A.C., acknowledges the support 
from St Catharine college in Cambridge and to the E. Oppeneheimer foundation.


\begin{thebibliography}{99}

\bibitem{Eley}  B. Alberts, D. Bray, A. Johnson, J. Lewis, M. Raff, K. Roberts, P. Walter 
``Essential Cell Biology'' Garland Publishing (1998)

\bibitem{Endres} R.G. Endres, D.L. Cox and R.R.P. Singh, Rev. Mod. Phys. {\bf 76}, 195 (2004) 

\bibitem{Seeman} N.C. Seeman, ``An overview of structural DNA nanotechnology'', Mol. Biotechnol. {\bf 37}, 246 (2007)

\bibitem{Rothemund} P.W.K. Rothemund, Nature {\bf 440}, 297 (16 March 2006)

\bibitem{YuHe} Yu He, Tao Ye, Min Su, Chuan Zhang, Alexander E. Ribbe, Wen Jiang and Chengde Mao, Nature {\bf 198}, 198 (2008)

\bibitem{Andersen} E.S. Andersen, M. Dong, M.M. Nielsen, K. Jahn, R. Subramani, 
W. Mamdouh, M.M. Golas, B. Sander, H. Stark, C.L.P. Oliveira, J.S. Pendersen, V. Birkedal, 
F. Besenbacher, K.V. Gothelf and J. Kjems, Nature {\bf 459}, 73 (2009)


\bibitem{fink} H.-W Fink and C. Sch\"onenberger, Nature {\bf 398} 407 (1999)

\bibitem{Barton1} M. R. Arkin, E. D. A. Stemp, R. E. Holmlin, J. K. Barton, A. Hoermann, E. J. C. Olson, and P. F. Barbara 
Science {\bf 273}, 475 (1996)

\bibitem{Barton2} D. B. Hall, R. E. Holmlin, and J. K. Barton, Nature, {\bf 382}, 731 (1996).

\bibitem{Porath} D. Porath, A. Bezryadin, S. de Vries and C. Dekker, Nature {\bf 403} 635 (2000)

\bibitem{Kasumov} A. Yu. Kasumov, M. Kociak, S. Gu\'eron,  B. Reulet, V. T. Volkov, D. V. Klinov and H. Bouchiat, 
Science {\bf 291} 280 (2001). 

\bibitem{Klinov} A. Yu. Kasumov, D. V. Klinov, P.-E. Roche, S. Gu\'eron, and H. Bouchiat, App. Phys. Let. {\bf 84} 1007 (2004);

\bibitem{DnaNewJourPhys} A.D. Chepelianskii, D. Klinov, A. Kasumov, S. Guéron, O. Pietrement, S. Lyonnais 
and H. Bouchiat, New J. Phys. {\bf 13}, 063046 (2011)


\bibitem{Pablo} P.J. de Pablo, F. Moreno-Herrero, J. Colchero, J. G\'omez Herrero, P. Herrero, A.M. Bar\'o, 
Pablo Ordej\'on, Jos\'e M. Soler and Emilio Artacho, PRL {\bf 85}, 4992 (2000) 

\bibitem{Storm} A. J. Storm, S. J. T. van Noort, S. de Vries, and C. Dekker, Appl. Phys. Lett. {\bf 79}, 3881 (2001)

\bibitem{zhangcox} Y. Zhang, R. H. Austin, J. Kraeft, E. C. Cox and N. P. Ong, PRL {\bf 89}, 198102 (2002)

\bibitem{Gomez} C. Gomez-Navarro, F. Moreno-Herrero, P. J. de Pablo, J. Colchero, J. Gomez-Herrero and A. M. Baro, PNAS {\bf 99}, 8484 (2002)
\bibitem{Heim1} T. Heim, D. Deresmes, and D. Vuillaume, Appl. Phys. Lett. {\bf 85}, 2637 (2004)

\bibitem{Heim2} T. Heim, D. Deresmes, and D. Vuillaume, J. Appl. Phys. {\bf 96}, 2927 (2004)

\bibitem{bingqian} Bingqian Xu, Peiming Zhang, Xiulan Li,and Nongjian Tao, Nano. Let. {\bf 4}, 1105 (2004)

\bibitem{porath} Errez Shapir, Hezy Cohen, Arrigo Calzolari, Carlo Cavazzoni, Dmitry A. Ryndyk, Gianaurelio Cuniberti, 
Alexander Kotlyar, Rosa Di Felice and  Danny Porath, Nature Materials {\bf 7}, 68 (2008) 

%
\bibitem{Barton} Xuefeng Guo, Alon A. Gorodetsky, James Hone, Jacqueline K. Barton and Colin Nuckolls, 
Nature Nanotechnology {\bf 3}, 163 (2008)


\bibitem{Stoeckenius} M.D. Stoeckenius, J. Biophys. and Biochem. Cytol., {\bf 11}, 297 (1961)

\bibitem{Hansma} P. K. Hansma, J. P. Cleveland, M. Radmacher, D. A. Walters, P. E. Hillner, M. Bezanilla, M. Fritz, D. Vie, H. G. Hansma, 
C. B. Prater, J. Massie, L. Fukunaga, J. Gurley, and V. Elings,  Appl. Phys. Lett. {\bf 64}, 1738 (1994)

\bibitem{ScienceAFM} L. Gross, F. Mohn,  N. Moll, P. Liljeroth, G. Meyer, Science {\bf 325} 1110 (2009)

\bibitem{Kawai} H. Tanaka and T. Kawai, Surface Science {\bf 539}, 531 (2003)

\bibitem{Pietrement} D. Pastr\'e, O. Pi\'etrement, S. Fusil, F. Landousy, J. Jeusset, M-O David, L. Hamon, E-L. Cam and A. Zozime 
Biophys. Jour. {\bf 85}, 2507 (2003)

\bibitem{Dykstra} {\it Biological Electron Microscopy},  Michael J. Dykstra and Laura E. Reuss, Springer (2003) ISBN: 978-0306477492

\bibitem{PNASab} J. M. Vargason, K. Henderson and P. Shing Ho, PNAS {\bf 98} 7265 (2001)

\bibitem{Croquette} J. F. Allemand, D. Bensimon,  R. Lavery and V. Croquette, PNAS {\bf 95} 14152 (1998)



\bibitem{pHens} J.F. Allemand, D. Bensimon, L. Jullien, A. Bensimon and V. Croquette 
Biophis. J. {\bf 73} 2064 (1997)


\bibitem{Dubochet} J. Dubochet, M. Ducommun, M. Zollinger and E. Kellenberger
 J. Ultrastruct. Res. {\bf 35} 147 (1971)

\bibitem{Liberman} M. A. Liberman and A. J. Lichtenberg, ``Principles of plasma discharges and materials processing'', A-Wiley interscience publication (1994)

\bibitem{Lide} D. R. Lide, ``Handbook of chemistry and physics'', CRC Press 84-th edition (2003), chapters 9-67 and 10-187



\bibitem{DnaPositive} A. Podest\`a, M. Indrieri, D. Brogioli, G. S. Manning, P. Milani, R. Guerra, L. Finzi and D. Dunlap, Biophysical Journal {\bf 89}, 2558 (2005)

\bibitem{DnaCombing} A. Bensimon, A. Simon, A. Chiffaudel, V. Croquette, F. Heslot and D. Bensimon, Science {\bf 265} 2096 (1994)

\bibitem{DNAflowENS} P. S. Doyle, B. Ladoux and J.-L. Viovy, Phys. Rev. Lett. {\bf 84} 4769 (2000)

\bibitem{Lionel} L. Angers, F. Chiodi, J. C. Cuevas, G. Montambaux, M. Ferrier, S. Gueron and H. Bouchiat, Phys. Rev. B {\bf 77}, 165408 (2008)

\bibitem{KasumovSWNT} A. Yu. Kasumov, R. Deblock, M. Kociak, B. Reulet, H. Bouchiat, I. I. Khodos, Yu. B. Gorbatov, V. T. Volkov, C. Journet and M. Burghard
Science {\bf 284}, 1508 (1999)

\bibitem{KasumovSWNT2} A. Kasumov, M. Kociak, M. Ferrier, R. Deblock, S. Gu\'eron, B. Reulet, I. Khodos, O. St\'ephan, and H. Bouchiat
Phys. Rev. B {\bf 68}, 214521 (2003)



\bibitem{Kasumov2} A.Yu. Kasumov, K. Tsukagoshi, M. Kawamura, T. Kobayashi, Y. Aoyagi, K. Senba, T. Kodama, H. Nishikawa, I. Ikemoto, 
K. Kikuchi, V.T. Volkov, Yu.A. Kasumov, R. Deblock, S. Gueron, and H. Bouchiat Phys. Rev. B {\bf 72}, 033414 (2005)

\bibitem{Kasumov3} A.Yu. Kasumov, S. Nakamae, M. Cazayous, T. Kawasaki, Y. Okahata
Research Letters in Nanotechnology, Article ID 540257 (2009)

\bibitem{Bartolo} R. E. Bartolo and N. Giordano, Phys. Rev. B {\bf 54} 3571 (1996)

\bibitem{Joyez} P. Joyez, D. Esteve and M.H. Devoret, Phys. Rev. Lett. {\bf 80}, 1956 (1998)

\bibitem{Kostia} M.V. Feigel'man, M.A. Skvortsov and K.S. Tikhonov, Solid State Comun. {\bf 149} 1101 (2009)

\bibitem{OrsayPhysics} Orsay physics, private communications 

\bibitem{PtSWNT} Jae-Hee Hana, Sun Hong Choia, Tae Young Leea, Ji-Beom Yoo, Chong-Yun Parka, Taewon Jungb, SeGi Yub, Whikun Yic, 
In Taek Hanb and Jong Min Kimb, Diamond and Related Materials {\bf 12}, 878 (2003)


\bibitem{Lvov} Yuri Lvov, Heinrich Haas, Gero Decher, Helmuth Moehwald and Alexei Kalachev J. Phys. Chem.,{\bf 97} 12835 (1993)

\bibitem{Rossnagel} S. M. Rossnagel, J.J. Cuomo and W.D. Westwood ``Handbook of plasma processing technology'', Noyes Publications, Park Ridge U.S.A (1990)

\bibitem{Nabok} A. Nabok ``Organic and inorganic nanostructures'', Artech House (2005) p.~39




\end{thebibliography}
\end{document}